\documentclass[bibyear]{aa}
\usepackage{graphicx}
\usepackage{txfonts}
\usepackage{breqn}
\usepackage{blindtext}
\begin{document}

\title{Seeking large-scale magnetic fields in a pure-disk dwarf galaxy NGC\,2976}

\author{
R. T. Drzazga\inst{1}
 \and K. T. Chy\.zy\inst{1}
 \and G. H. Heald\inst{2,3*}
 \and D. Elstner\inst{4}
 \and J. S. Gallagher III\inst{5}
}
\institute{Obserwatorium Astronomiczne Uniwersytetu
Jagiello\'nskiego, ul. Orla 171, 30-244 Krak\'ow, Poland
\and ASTRON, the Netherlands Institute for Radio Astronomy, Postbus 2, 7990 AA, Dwingeloo, The Netherlands
\and Kapteyn Astronomical Institute, University of Groningen, PO Box 800, 9700 AV, Groningen, The Netherlands
\and Leibniz Institute for Astrophysics Potsdam, An der Sternwarte 16, 14482 Potsdam, Germany
\and Department of Astronomy, University of Wisconsin-Madison, 5534 Sterling, 475 North Charter St., Madison WI 53706, USA
}
\offprints{R.T. Drzazga}
\mail{drzazga@oa.uj.edu.pl \\
*\,Current address: CSIRO Astronomy and Space Science, 26 Dick Perry Avenue, Kensington, Perth WA 6151, Australia}
\date{Received date/ Accepted date}
\titlerunning{Seeking large-scale magnetic fields in a pure-disk dwarf galaxy NGC\,2976}
\authorrunning{R.T. Drzazga et al.}

\abstract
{}
{It is still  unknown how  magnetic field-generation mechanisms could operate in low-mass dwarf galaxies. Here, we present a 
detailed study of a nearby pure-disk dwarf galaxy NGC\,2976. Unlike  previously observed dwarf objects, this galaxy 
possesses a clearly defined disk. We also discuss whether NGC\,2976 could serve as a potential source of the intergalactic magnetic field.}
{For the purpose of our studies, we performed deep multi-frequency polarimetric observations of NGC\,2976 with 
the VLA and Effelsberg radio telescopes. Additionally, we supplement them with re-imaged data from the WSRT-SINGS survey for 
which a rotation measure (RM) synthesis was performed.
A new weighting scheme for the RM synthesis algorithm, consisting of including information about the quality 
of data in individual frequency channels, was proposed and investigated. Application of this new weighting to 
the simulated data, as well as to the observed data, results in an improvement of the signal-to-noise ratio in 
the Faraday depth space.}
{The magnetic field morphology discovered in NGC\,2976 consists of a southern polarized ridge. This structure 
does not seem to be due to just a pure large-scale dynamo process (possibly cosmic-ray driven) 
at work in this object, as indicated by the RM data and dynamo number calculations. 
Instead, the field of NGC\,2976 is modified by past gravitational interactions and possibly also 
by ram pressure inside the M\,81 galaxy group environment. The estimates of total (7\,$\mu$G) and 
ordered (3\,$\mu$G) magnetic field strengths, as well as degree of field order (0.46), which is similar to those observed in spirals, 
suggest that tidally generated magnetized gas flows can further enhance dynamo action in the object.
NGC\,2976 is apparently a good candidate for the efficient magnetization of its neighbourhood. It is able to provide an ordered 
(perhaps also regular) magnetic field into the intergalactic space up to a distance of about 5\,kpc.}
{Tidal interactions (and possibly also ram pressure) can lead to the formation of unusual magnetic field morphologies
(like polarized ridges) in galaxies  out of the star-forming disks, which do not follow any observed component of the interstellar medium (ISM), 
as observed in NGC\,2976. These galaxies are able to provide ordered magnetic fields far out of their main disks.}

\maketitle

\section{Introduction}
\label{s:n2976Introduction}
In low-mass dwarf galaxies, the conditions and efficiency of the magnetic 
field generation process are still not very well understood (see e.g. Chy\.zy et~al. \cite{chyzy11}). 
In these objects, the influence of magnetic fields on the interstellar medium is expected to be even more 
important, because of their smaller gravitational potential and the possibility of 
gas escaping via galactic winds. On the other hand,  dwarf galaxies rotate more 
weakly  with more random motions, providing for a weak $\omega$ -- effect.
Therefore, there could be no large-scale dynamo at work in some dwarfs at all, 
as they are below the dynamo efficiency threshold (Gressel et~al. \cite{gressel08}). In these objects, random magnetic 
fields could be produced by a small-scale dynamo action (Zeldovich et~al. \cite{zeldovich90}).

The knowledge of properties of magnetic fields in dwarfs is also crucial for explaining the magnetization of the intergalactic medium (IGM). 
Kronberg et~al. (\cite{kronberg99}) suggested in their `boiling universe' concept that  low-mass galaxies were likely sources of primeval fields. 
However, Chy\.zy et~al. (\cite{chyzy11}) argued that for a sample of dwarf galaxies from the Local Group, these objects seem to be only able to magnetize their immediate neighbourhood. A similar conclusion was also drawn by Drzazga et~al. (\cite{drzazga11}) in their 
studies of interacting galaxies as potential sources of the IGM's magnetic fields. Thus the question about the origin of the primeval 
magnetic fields and the role of dwarf galaxies in the magnetization of the Universe remains  unanswered.

Only a handful of dwarf galaxies have been observed to date with respect to magnetic 
fields for which the dynamo concepts could be tested. As expected, in the very small 
and low-mass dwarfs IC\,10 and NGC\,6822 (linear size of 1 and 2 kpc, respectively) no 
large-scale (regular) magnetic fields have been detected to date (Chy\.zy et~al. \cite{chyzy03}, 
Chy\.zy et~al. \cite{chyzy11}, Heesen et~al. \cite{heesen11}), 
which indicates that these objects are below the efficiency threshold of the large-scale dynamo.
In the more massive and larger dwarfs, NGC\,4449 as well as in the Large Magellanic 
Cloud (LMC; of 7 and 10\,kpc size, respectively), regular magnetic fields (of 8 and 
1 $\mu$G, respectively) with fragments of some large-scale spiral structure, as evident 
signatures of large-scale $\alpha - \omega$ dynamo, were observed (Chy\.zy et~al. \cite{chyzy00}, 
Gaensler et~al. \cite{gaensler05}). Some regular fields (of about 1 $\mu$G) were also found in 
the Small Magellanic Cloud (SMC; of 6 kpc size), but in this case, the field structure 
did not resemble any large-scale dynamo (Mao et~al. \cite{mao08}). However, as these 
three large dwarfs show distinct signs of strong tidal interactions, it is not 
clear whether and how gravitational interactions influenced the set-up of the 
large-scale dynamo process, and which conditions are appropriate for an efficient 
dynamo to occur in dwarfs.

To shed new light on the problem of the generation and evolution of magnetic fields in low-mass galaxies, multi-frequency, 
sensitive radio polarimetric observations of a dwarf galaxy NGC\,2976 were performed. This is a dynamically simple, 
bulgeless, pure-disk object, with no discernible spiral arms 
(Simon et~al. \cite{simon03}) 
of SAc type (De Vaucouleurs et~al. \cite{deVaucouleurs91}) located at a distance of 3.6\,Mpc. 
This galaxy seems to be a scaled version of the class of larger pure-disk spirals 
(e.g. Gallagher \& Matthews \cite{gallager02}, Matthews \& van Driel \cite{matthews00}). 
It has an absolute visual luminosity and disk size between those of SMC and LMC.
The unique characteristics of NGC\,2976 -- its morphology of a typical 
spiral galaxy but its linear size and mass corresponding to those of dwarfs -- makes it an ideal 
laboratory to study magnetic field-generation processes in low-mass galaxies.
The total power (TP) radio emission of NGC\,2976 was observed at 1.4\,GHz with 
the Very Large Array (VLA) by Condon (\cite{condon87}). 
This galaxy was also included in the WSRT-SINGS polarimetric survey (Heald et~al. \cite{heald09}; Braun et~al. \cite{braun10}), within 
which deep and high-resolution observations were performed using the WSRT. 
However, the relatively small mass and the dwarf-like, rather than spiral character of NGC\,2976 ,
went unnoticed in  previous studies.  

NGC\,2976 is a member of the M\,81 galaxy group, located close to its centre.
However, unlike the Magellanic Clouds and NGC\,4449, NGC\,2976 has probably not been  greatly 
disturbed for a long time, as it shows very regular U, B, and V isophotes in the outer disk 
(Bronkalla et~al. \cite{bronkalla92}), its \ion{H}{i} velocity field is undistorted, and its velocity dispersion 
is small (11\,km s$^{-1}$, Stil \& Israel \cite{stil02b}). Hence, the conditions for the magnetic 
field generation process (e.g. galactic dynamics, size, mass, star formation rate (SFR), etc.) 
can be constrained for this object without the strong influence of gravitational or ram pressure interactions. 
Such well-established dynamo conditions are of primary significance in building up 
theoretical concepts of the dynamo theory (Brandenburg \& Subramanian \cite{brandenburg05}) as 
well as in explaining the magnetic structures that have been observed in dwarfs.

\begin{table*}
\caption{Basic properties of NGC\,2976}
\begin{center}
\begin{tabular}{lccccc}
\hline\hline
Name             & Hubble $^{(1)}$ & Inclination $^{(1)}$ & Linear size $^{(2)}$  & \ion{H}{i} mass$^{(3)}$  & SFR $^{(3)}$          \\
                 & type            & [deg.]               & [kpc]                 & [10$^8$ M$_\odot$] & [M$_\odot$ / yr]      \\
\hline
NGC\,2976  &  SAc      & 65            &  6            & 1.8 (1.5$^{(4)}$) & 0.2     \\
\hline
\end{tabular}
\end{center}
{\bf References.} $1$ - Heald et~al. (\cite{heald09}), $2$ - estimates based on distances taken from Kennicutt et~al. (\cite{kennicutt03}) and angular sizes (D$_{25}$) from 
Heald et~al. (\cite{heald09}), $3$ - Kennicutt et~al. (\cite{kennicutt03}), $4$ - Stil \& Israel (\cite{stil02b})
\label{t:wsrtSingsIntroTab}
\end{table*}

Studies of magnetic fields in dwarf galaxies are observationally difficult, as these objects are usually radio-weak (e.g. Chy\.zy et~al. \cite{chyzy11}). 
Hence, we used sensitive instruments: the 
VLA\footnote{The National Radio Astronomy Observatory is a facility of the National Science Foundation operated under cooperative agreement by Associated Universities, Inc.} 
(USA) and 100-meter Effelsberg 
radio telescope\footnote{The 100-m telescope at Effelsberg is operated by the Max-Planck-Institut f\"ur Radioastronomie (MPIfR) on behalf of the Max-Planck-Gesellschaft.} 
(Germany) to observe NGC\,2976 . Additionally, we supplemented our observations with the spectro-polarimetric data from the WSRT-SINGS survey (Heald et~al. \cite{heald09}). 
The detailed account of these observations and the data reduction are given in Section 2. In Section 3 the results are presented. 
Discussion and conclusions of the results are shown in Sections 4 and 5, respectively. 

\section{Observations and data reduction}
\label{s:n2976Observations} 
\subsection{VLA}
\label{ss:n2976VLAReduction}

NGC\,2976 was observed with the VLA in L band in C, DnC, and D configurations (Table \ref{t:n2976vla}). The total on-source time, as 
calculated from the schedule and dynamic times, was about 16 hours. The observations were performed in  full polarization mode using 
two 50\,MHz width `intermediate frequency' channels centred at 1385.100 and 1464.900\,MHz. The data were reduced with  the AIPS package\footnote{http://www.aips.nrao.edu/index.shtml}, 
following the standard procedure. The flux scale and position angle of polarization were calibrated using the 
amplitude calibrator (0521+166;3C\,138/1331+305;3C\,286). 
The phase calibrator (0841+708) was used to find gain, phase, and polarization leakage solutions. Each of the datasets specified in 
Table \ref{t:n2976vla} was calibrated in this way separately, and then self-calibrated in phase-only. 
After checking for consistency between the datasets, they were concatenated and self-calibrated in phase, as well as in amplitude and phase. 

In the final stage, imaging with the AIPS task IMAGR was performed for all the Stokes parameters using Briggs' robust weighting 
of 1 and -5. The obtained Q and U maps were combined within the task COMB to form maps of 
polarized intensity that were corrected for positive bias ($PI = \sqrt{Q^2 + U^2}$) and 
polarization angle ($PA = 0.5 \arctan(U/Q)$). The rms sensitivity (for imaging with robust 1) is 20\,$\mu$Jy/beam for the total power 
intensity, and 9\,$\mu$Jy/beam for the polarized intensity (PI).

\begin{table*}
\caption{Parameters of the VLA observations of NGC\,2976 at 1.43\,GHz.}
\begin{center}
\begin{tabular}{cccccc}
\hline
\hline
Type of obs.& Obs. date   & Configuration & Integr. time & Amp. cal.         & Phas. cal. \\
            &             &               & [h]          &                   &            \\
\hline
   Schedule & 28 Aug 2009 & C             & 5.8          & 0521+166/1331+305 & 0841+708   \\
   Dynamic  & 18 Sep 2009 & DnC           & 0.7          & 0521+166          & 0841+708   \\
   Dynamic  & 20 Sep 2009 & DnC           & 1.4          & 0521+166          & 0841+708   \\
   Dynamic  & 21 Sep 2009 & DnC           & 1.6          & 0521+166          & 0841+708   \\
   Dynamic  & 01 Oct 2009 & DnC           & 1.6          & 0521+166          & 0841+708   \\
   Schedule & 31 Oct 2009 & D             & 1.5          & 1331+305          & 0841+708   \\
   Dynamic  & 06 Nov 2009 & D             & 1.6          & 1331+305          & 0841+708   \\
   Dynamic  & 11 Nov 2009 & D             & 1.6          & 0521+166          & 0841+708   \\
\hline
\end{tabular}
\end{center}
\label{t:n2976vla}
\end{table*}

\subsection{Effelsberg}
\label{ss:n2976EffelsbergReduction}

NGC\,2976 was observed with the 100-m Effelsberg radio telescope at 4.85 and 8.35\,GHz. At 4.85\,GHz, a two-horn secondary focus system with bandwidth of 
0.5\,GHz was used, recording data in four channels/horn. Two of these channels recorded the total power signal, while the others recorded signals 
in Stokes Q and U. For the object, 24 coverages (maps) in Az/Alt frame were obtained. However, only 12 of them were suitable for polarization studies 
(Table \ref{t:n2976eff}). At 8.35\,GHz, a 1.1\,GHz bandwidth single-horn receiver was used, also recording data in four channels (see above). In this 
case, 66 coverages in the right ascension-declination (RA/Dec) frame were obtained. Unfortunately, nearly half of the data (Table \ref{t:n2976eff}) 
were unusable, owing to strong radio frequency interference-like (RFI) spikes near the target. At both the observation frequencies, corrections of telescope pointing were made about every 1.5 hours by scanning a 
nearby strong point radio source. 

We performed data reduction in the standard way by applying the NOD2 package (Haslam \cite{haslam74}). First, all coverages at both the frequencies were 
edited to remove any RFI and align baselines to the common level; then the data at 4.85\,GHz from both horns were averaged using the 
software beam-switching technique (Morsi \& Reich \cite{morsi86}) and transformed to the RA/Dec frame (for 8.35\,GHz data this step was not necessary, see 
above). In the next stage of data reduction, all usable coverages (Table \ref{t:n2976eff}) were combined using the spatial frequency weighting 
method (Emerson \& Gr\"ave \cite{emerson88}). In the final stage, all Stokes parameter maps at both wavelengths were digitally filtered to 
remove Fourier spatial frequencies, which correspond to noisy structures that are less than the beam sizes and then transformed to the FITS format. 
Further processing that included flux scale calibration and the formation 
of polarized intensity (corrected for positive bias) and polarization angle maps was performed in the AIPS package. 
At both  frequencies, the flux scales were calibrated using 3C\,286, based on the assumption that 
its flux is 7.47\,Jy at 4.85\,GHz and 5.27\,Jy at 8.35\,GHz (as provided in the VLA Calibrator 
List\footnote{https://science.nrao.edu/facilities/vla/observing/callist}). The obtained rms noise levels are given in Table \ref{t:n2976eff}.

\begin{table*}
\caption{Parameters of the Effelsberg observations of NGC\,2976 at 4.85 and 8.35\,GHz}
\begin{center}
\begin{tabular}{ccccccccc}
\hline
\hline
Obs. freq.   & Obs. date & Map size       & Total/good(TP)/good(PI)       & S$_{TP}$            & S$_{PI}$            & r.m.s. (TP)    & r.m.s. (PI) \\
 GHz         &           & arcmin         & coverages                     &   mJy             &    mJy            &     mJy/beam & mJy/beam  \\
\hline
   4.85      & May 2009          & 35 $\times$ 25 &  24/21/12                        & 30.0 $\pm$ 1.4      & 1.6 $\pm$ 0.4       & 0.45           & 0.08        \\
             & May 2010          &                &                                  &                     &                     &                &             \\
             & May 2011          &                &                                  &                     &                     &                &             \\
   8.35      & May 2011          & 21 $\times$ 21 &  66/35/35                        & 19.3 $\pm$ 0.9      & n.a.          & 0.23           & 0.08        \\
\hline
\end{tabular}
\end{center}
\label{t:n2976eff}
\end{table*}

\subsection{Re-analysis of the WSRT-SINGS data}
\label{ss:WsrtSingsReduction}

We also made use of the calibrated WSRT-SINGS polarimetric uv data in the 18\,cm and 22\,cm bands 
(for details see Heald et~al. \cite{heald09}). The data were re-imaged by applying the Miriad software using  the Briggs robust 
weighting of 1.0. The Q and U frequency data cubes containing 702 frequency channels obtained in this way were convolved to the 
resolution of 48.6$"$ $\times$ 45.0$"$, providing a high sensitivity to extended radio structures.
Then, for the Q and U data cubes, rotation measure synthesis (RM synthesis; Brentjens \& de Bruyn \cite{brentjensAndDeBruyn05}) was 
performed applying a new weighting scheme that takes the quality of individual frequency channels into account (see Appendix \ref{ch:newWeighting} for details).

The resolution in Faraday depth that was obtained for the frequency coverage of the WSRT-SINGS survey is about 144\,rad m$^{-2}$, 
whereas the sensitivity to extended structures in Faraday space (at which 50\% of polarized flux can be recovered) is about 
110\,rad m$^{-2}$. The maximum Faraday depth that could be sampled in the Faraday domain is about 1.7 $\times$ 10$^5$ rad m$^{-2}$. 
We did not expect polarized emission at such high depths so, the Faraday domain was sampled within a range of -500 to +500\,rad m$^{-2}$ every 
2.5\,rad m$^{-2}$. As the RMTF (rotation measure transfer function) shows a high level of side-lobes 
(of about 80\%, see Sect. \ref{s:strongAndWeakSideOfWeighting}) owing to the gap in frequency coverage between 18\,cm and 22\,cm data bands, 
the RM-CLEAN deconvolution method (Heald et~al. \cite{heald09}) was used. With this algorithm, the Q($\phi$) and U($\phi$) cubes 
were cleaned down to the 1-sigma noise level and then restored with a Gaussian beam of the same width as the main RMTF lobe. 
After cleaning Q and U data cubes, we calculated the polarized intensity map for the studied object 
(not corrected for the positive bias as negligible in this case). This map was obtained as a maximum of PI signal 
over all sampled Faraday depths for each line of sight. We also computed B-vectors, maximum Faraday depth (for the maximum of PI signal), 
and magnetic field vector distribution maps. The B-vectors map was calculated for the effective frequency of about 1.5\,GHz to which all 
polarization vectors were de-rotated. This approach is applied because, for the given resolution in Faraday domain and signal-to-noise ratio 
of at least 4, the maximum Faraday depth error is large, about 18\,rad m$^{-2}$, which prevents us from directly obtaining  reliable 
information about intrinsic directions of magnetic field vectors (see Brentjens \& de Bruyn \cite{brentjensAndDeBruyn05}).

\section{Results}
\label{s:n2976Results}

\subsection{1.43\,GHz}
\label{ss:n2976_1_4GHz}
\begin{figure}
\centering
\includegraphics[width=0.5\textwidth,clip=true]{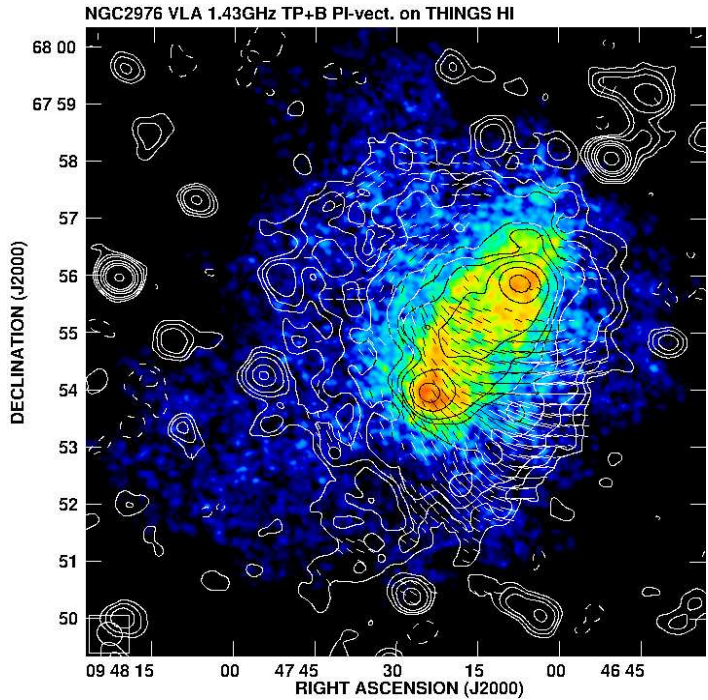}
\caption{Total-power contours and apparent B-vectors (not corrected for Faraday rotation) of polarized intensity of NGC\,2976 (obtained with Robust=1 weighting) 
at 1.43\,GHz superimposed on the \ion{H}{i} map (from the THINGS survey, Walter et~al. \cite{walter08}) shown in a logarithmic scale. 
The \ion{H}{i} map resolution is 7.41$"$ $\times$ 6.42$"$ HPBW, position angle of the beam is 71.8 degrees.
The contours levels are (-5, -3, 3, 5, 8, 16, 24, 32, 64, 128, 256) $\times$ 20 (rms noise level) $\mu$Jy/beam. A vector of 10" length corresponds 
to the polarized intensity of about 42\,$\mu$Jy/beam. The map resolution is 26.8$"$ $\times$ 23.5$"$ HPBW, while the beam position angle is 62 degrees.}
\label{f:n2976_1_4_tpHI}
\end{figure}

\begin{figure}
\centering
\includegraphics[width=0.49\textwidth,clip=true]{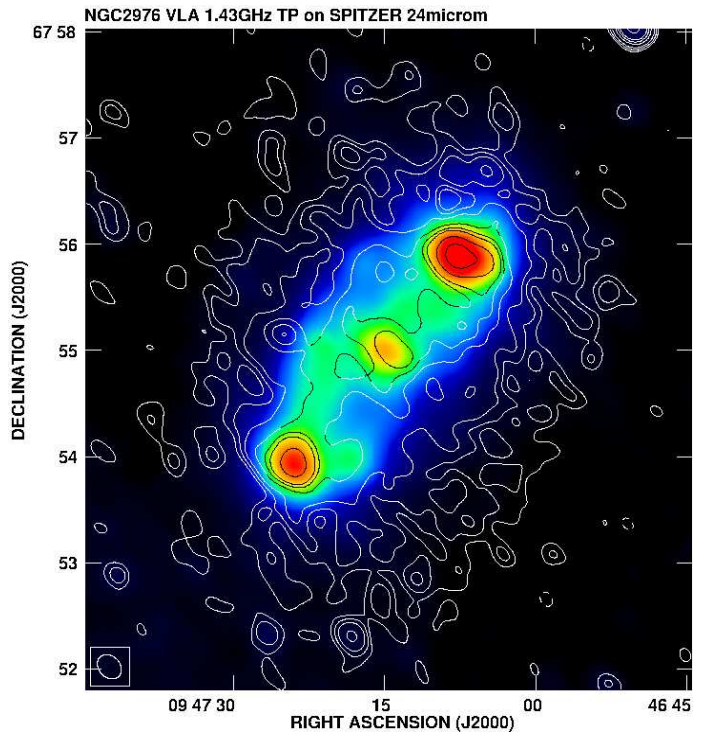}
\caption{Total-power contours of NGC\,2976 (obtained with Robust=-5 weighting) at 1.43\,GHz superimposed on the 
{\it Spitzer} 24\,$\mu$m image (Kennicutt et~al. \cite{kennicutt03}), shown in logarithmic scale and convolved to the resolution of radio data. 
The contours levels are (-5, -3, 3, 5, 8, 16, 24, 32, 64, 128) $\times$ 20 (rms noise level) $\mu$Jy/beam. 
The map resolution is 15.1$"$ $\times$ 11.9$"$ HPBW, while the beam position angle is 42 degrees.}
\label{f:n2976_1_4_tp24}
\end{figure}

Our sensitive polarimetric VLA observation of NGC\,2976 results in a detection of some very extended radio emission surrounding the object 
(Fig. \ref{f:n2976_1_4_tpHI}). This radio extent of the galaxy was not known earlier, as previously only a total power emission closely related to 
its optical disk was obtained (Condon \cite{condon87}; Niklas \cite{niklas95}; Stil \& Israel \cite{stil02a}; Heald et~al. \cite{heald09}). 
In the low-resolution total power map (Fig. \ref{f:n2976_1_4_tpHI}), the brightest emission coincides with the sharpened disk 
(i.e. a disk possessing sharp truncation edge; Williams et~al. \cite{williams10}) of the NGC\,2976, and the maximum is well correlated 
with two bright \ion{H}{ii} regions located at the edges of the object. This is even more visible in the high-resolution (of about 0.26\,kpc) map 
(Fig. \ref{f:n2976_1_4_tp24}) superimposed on the 24\,$\mu$m infrared image from {\it Spitzer} (Kennicutt et~al. \cite{kennicutt03}). There is a 
good correspondence between the images in both wavelengths.

Beside the stellar disk, the total power radio emission (in the lower resolution map, Fig. \ref{f:n2976_1_4_tpHI}) extends in the 
north-eastern direction, where a low-brightness signal of about 60\,$\mu$Jy/beam (3 $\times$ rms noise level) resides. Radio emission at a 
similar level is also observed in the southern part of the galaxy. However, here its extension is smaller and brighter than in the opposite 
direction. The total power radiation of NGC\,2976 corresponds well to its \ion{H}{i} distribution 
(Fig. \ref{f:n2976_1_4_tpHI}) in the NE and SW directions. However, radio emission does not follow the outer SE extension  
and the spur of \ion{H}{i} in the north (see also Chynoweth et~al. \cite{chynoweth08}).

\begin{figure}
\centering
\includegraphics[width=0.49\textwidth,clip=true]{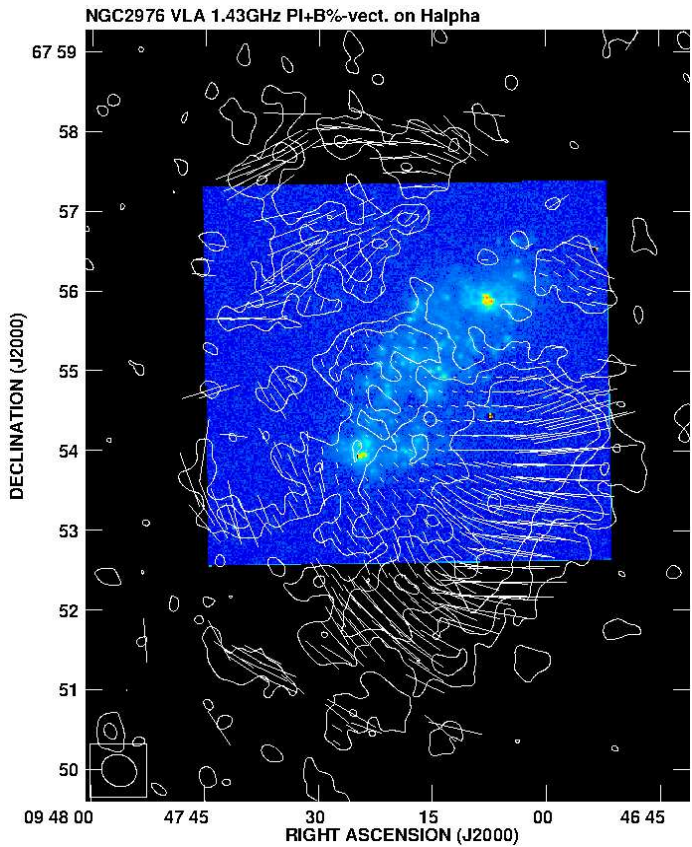}
\caption{Contours of polarized intensity and apparent B-vectors (not corrected for Faraday rotation) of polarization degree of NGC\,2976  
(obtained with Robust=1 weighting) at 1.43\,GHz superimposed on the H$\alpha$ image (Dale et~al. \cite{dale09}). The contour levels are 
(3, 5, 8) $\times$ 9 (rms noise level) $\mu$Jy/beam. A vector of 10" length corresponds to the polarization degree of 12.5\%. The map resolution is 
26.8$"$ $\times$ 23.5$"$ HPBW, while the beam position angle is 62 degrees.}
\label{f:n2976_1_4_piHalpha}
\end{figure}

\begin{figure}
\centering
\includegraphics[width=0.49\textwidth,clip=true]{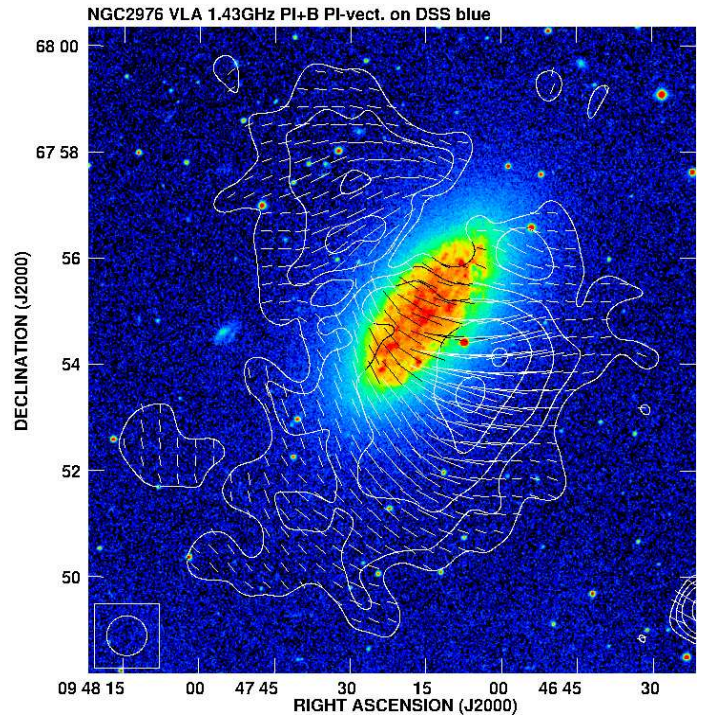}
\caption{Contours and apparent B-vectors (not corrected for Faraday rotation) of the polarized intensity of NGC\,2976 at 1.43\,GHz convolved to the resolution of 45$" $ and superimposed on 
the DSS blue image. The contour levels are (3, 5, 8, 16, 24) $\times$ 13 (rms noise level) $\mu$Jy/beam. 
The vector of 10" length corresponds to the polarized intensity of about 42\,$\mu$Jy/beam.}
\label{f:n2976_1_4_pi45}
\end{figure}

The distribution of polarized intensity in NGC\,2976 (Fig. \ref{f:n2976_1_4_piHalpha}) is even more surprising than the total power one. 
Most of the observed polarized emission (more than 60\%) comes from the south-western part of the object, where the polarization degree 
reaches 50\%. We note that this number does not take into account the regions within the object close to the total power 
intensity at the level of 3$\times$rms, where the polarization degree has unphysical values as high as 100\%. This is due to 
the higher sensitivity of the polarized intensity map, which is not bounded by the confusion limit. The brightest southern PI region 
was also detected by Heald et~al. (\cite{heald09}) in their high-resolution WSRT-SINGS survey 
(see also Sect. \ref{ss:n2976RMSynthesisResults}). The polarized signal is weaker in the disk, and appears only in its southern part. 
In the northern part of NGC\,2976, the distribution of polarized signal forms a curved arc structure. A convolution of 
the Q and U Stokes parameters maps to the lower resolution of 45$"$ and then computing the PI map that resulted from them (Fig. \ref{f:n2976_1_4_pi45}) 
enabled us to also detect  a polarized extension in the southern part of the object. Southern and south-eastern extensions coincids 
with \ion{the H}{i} tail and the northern tip of PI corresponds to the northern \ion{H}{i} spur, which is visible in Fig. \ref{f:n2976_1_4_tpHI}. 

Comparison of the polarized intensity map of NGC\,2976 with its H$\alpha$ emission (Fig. \ref{f:n2976_1_4_piHalpha}) reveals that the PI 
distribution of the object is quite unusual  compared to the other low-mass galaxies for which polarimetric observations were performed. Typically 
for dwarfs, the polarized signal is closely related to the regions where massive stars are formed. These findings were made in NGC\,4449, where the 
emission follows a part of the massive arm that  is actively forming stars (Chy\.zy et~al. \cite{chyzy00}); IC10, in which only a polarized intensity near the giant \ion{H}{ii} 
regions was detected (Chy\.zy et~al. \cite{chyzy03}; Heesen et~al. \cite{heesen11}); NGC\,1569, where multi-wavelength radio observations reveal PI features 
that are associated with H$\alpha$ bubbles -- outside the stellar disk (Kepley et~al. \cite{kepley10}). However, in NGC\,2976 the polarized emission mostly resides  very 
far from \ion{H}{ii} regions. This is likely due to external agents (tidal interactions and perhaps ram pressure, 
see Sect. \ref{ss:n2976TidalvsRamPressure}) and is, to some degree, also due to the Faraday depolarization. 
Faint H$\alpha$ emission extends only to the west from the northern \ion{H}{ii} region and to the east from the central part of the system.
Even trying to increase the signal-to-noise ratio in the H$\alpha$ map by convolving it to 
the resolution of radio data does not reveal any extended emission\footnote{The same is true for the map in far-UV, which is also a tracer of star 
formation.}.  This is in agreement with Thilker et~al. (\cite{thilker07}), where the galaxy was classified as not having an extended UV disk. 
While NGC\,2976 shares a high level of optical symmetry with M\,33, which also has a thin disk and a rotation that is only twice as fast (Tabatabaei et~al. 
\cite{tabatabaei08}), the polarized emission of NGC\,2976 is much more asymmetric.

Based on the WSRT-SINGS survey, Heald et~al. (\cite{heald09}) found that, in normal galaxies, their receding side is significantly depolarized 
at long wavelengths (of about 20\,cm), owing to a geometry of large-scale magnetic fields and the longer way the polarized waves travel through the 
magneto-ionic medium. From the \ion{H}{i} velocity field observations of NGC\,2976 (publicly available within the 
THINGS survey, Walter et~al. \cite{walter08}) we found that the north-western part of the object has a receding major axis. 
This implies that this side of the galaxy should exhibit a less polarized emission,  compared to the opposite side. 
This effect is actually observed for the optical disk.

The total power flux density measured at 1.43\,GHz for NGC\,2976 is 65.7 $\pm$ 2.8\,mJy, while the polarized intensity flux 
density is 7.8 $\pm$ 0.8\,mJy, which gives a total polarization degree of 12 $\pm$ 2\%. Such values are typically observed 
in more massive spiral galaxies but at higher radio frequencies (e.g. Beck et~al. \cite{beck96}). 

\subsection{4.85 and 8.35\,GHz}

\begin{figure*}
\centering
\includegraphics[width=0.39\textwidth,clip=true]{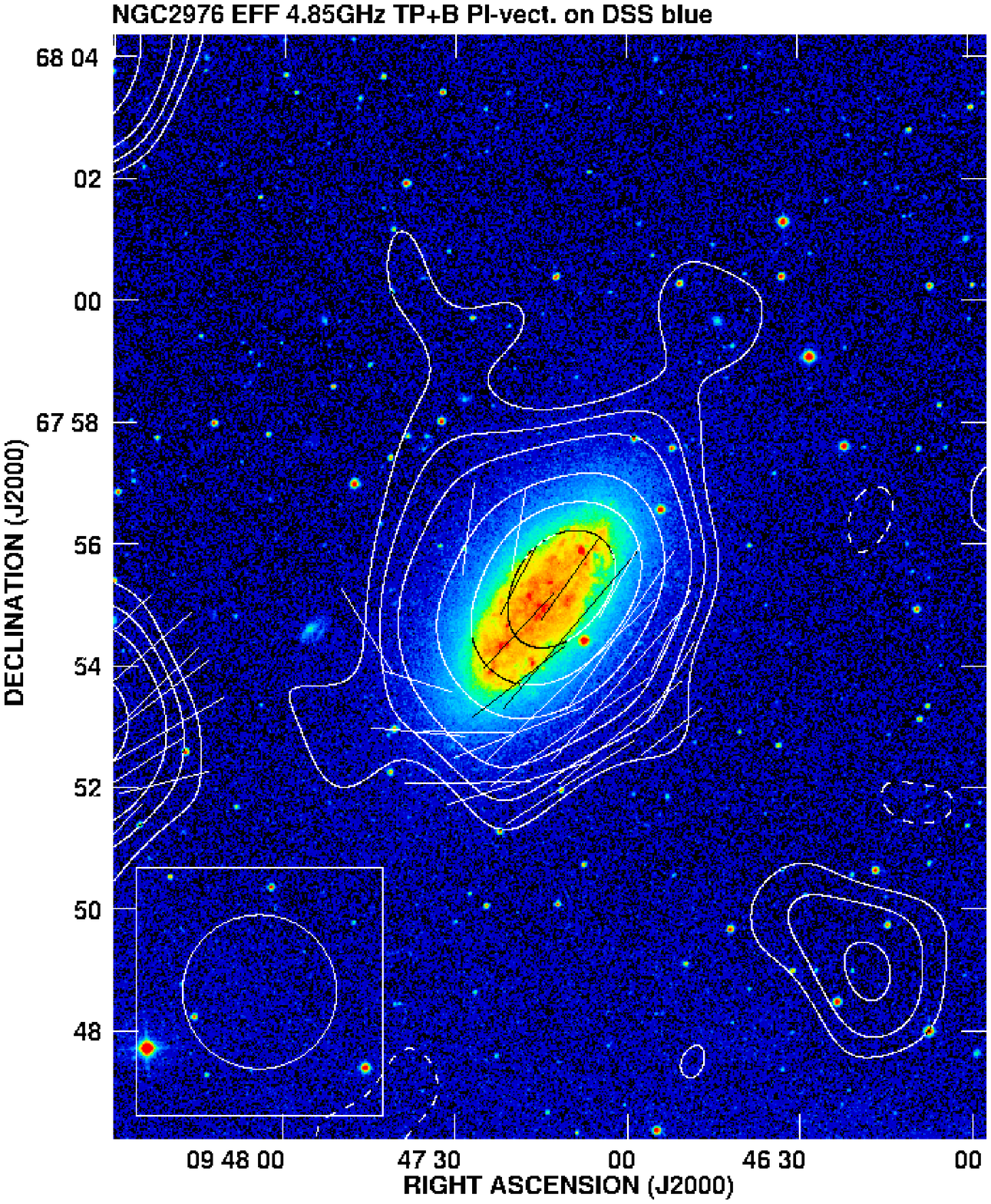}
\includegraphics[width=0.39\textwidth,clip=true]{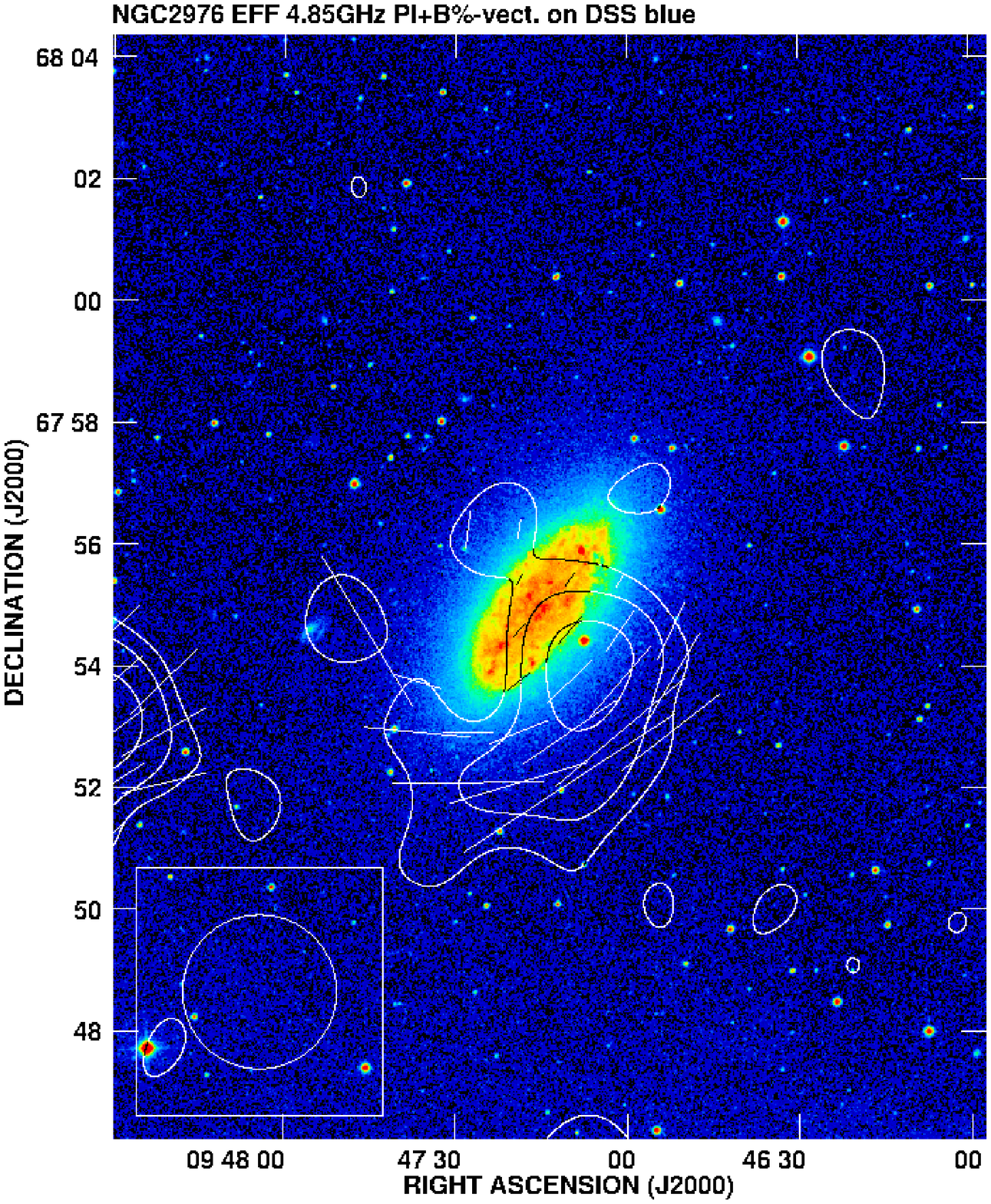}
\caption{Left: Total-power contours and apparent B-vectors (not corrected for Faraday rotation) of polarized intensity of NGC\,2976 at 4.85\,GHz 
superimposed on the DSS blue image. The contours levels are (-5, -3, 3, 5, 8, 16, 24, 32) $\times$ 450 (rms noise level) $\mu$Jy/beam.
The vector of 10" length corresponds to the polarized intensity of 33\,$\mu$Jy/beam. 
Right: Contours of polarized intensity and apparent B-vectors (not corrected for Faraday rotation) of polarization degree of NGC\,2976 at 4.85\,GHz 
superimposed on the DSS blue image. The contour levels are (3, 5, 8) $\times$ 80 (rms noise level) $\mu$Jy/beam. The vector of 10" length 
corresponds to the polarization degree of 1.3\%. The maps resolution is 152$"$ $\times$ 152$"$ HPBW.}
\label{f:n2976_6cm}
\end{figure*}

In Figure \ref{f:n2976_6cm}, the total power emission at 4.85\,GHz from the Effelsberg telescope is presented (see also Jurusik et~al. \cite{jurusik14}). 
Its resolution is much lower (152$"$) than in the VLA observations. In this map, similarly to the high-resolution data, the TP signal has a 
maximum shifted to the northern part of the galaxy's disk, where one of the giant \ion{H}{ii} regions is located. The low-level 
emission of 1.35 - 2.25\,mJy/beam is elongated in the NE direction (possibly it is not real). 
There is no such elongation to SE. A similar asymmetry was observed in the 1.43\,GHz data (Fig. \ref{f:n2976_1_4_tpHI}). 
The elongation in the NW direction seems to be correlated with a source that is also visible  at 1.43\,GHz. The total power flux density of NGC\,2976 measured 
at 4.85\,GHz is 30.0 $\pm$ 1.4\,mJy (Table \ref{t:n2976eff}), which is in excellent agreement with the radio flux that is determined at 
the same frequency by Gregory \& Condon (\cite{gregory91}).

The morphology of polarized intensity at 4.85\,GHz (Fig. \ref{f:n2976_6cm}) closely resembles that at 1.43\,GHz. 
Here almost all  (about 80\%) of the polarized flux density, which is 1.6 $\pm$ 0.4\,mJy (giving a polarization degree of about 5\%, Table \ref{t:n2976eff}), 
comes from the southern part of the disk and the southern extension. Only some small polarized blobs have been detected in northern direction. 
The narrow emission from the disk is located close to the galaxy centre, while at 1.43\,GHz it is barely shifted towards the south.

The noticeable difference in the polarized emission between 4.85\,GHz and 1.43\,GHz is related to the orientation of the B-vectors. At 4.85\,GHz, the vectors 
are nearly perpendicular to those observed at the lower frequency. As the Faraday effects are smaller at shorter wavelengths, we can expect that the 
directions of B-vectors, which are observed at 4.85\,GHz, are close to the intrinsic directions of magnetic field vectors (see Fig. \ref{f:N2976AndSingsBfields}). 

\begin{figure}
\centering
\includegraphics[width=0.39\textwidth,clip=true]{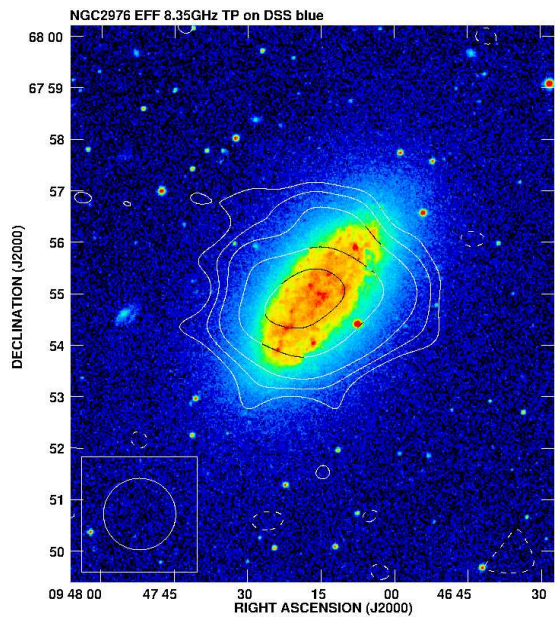}
\caption{Total-power contours of NGC\,2976 at 8.35\,GHz 
superimposed on the DSS blue image. The contours levels are (-5, -3, 3, 5, 8, 16, 24) $\times$ 230 (rms noise level) $\mu$Jy/beam.
The map resolution is 84$"$ $\times$ 84$"$ HPBW.}
\label{f:n2976_3cm}
\end{figure}

The total power emission of NGC\,2976 at 8.35\,GHz is presented in Fig. \ref{f:n2976_3cm}. This is closely related to the optical disk. There is no 
radio emission extension to the NE direction seen at lower frequencies, as expected for steep-spectrum radiation (see Sect. \ref{ss:n2976SpectralIndex}).
The total power flux density measured for this frequency is 19.3 $\pm$ 0.9\,mJy (Table \ref{t:n2976eff}). In the polarization, we 
did not detect the galaxy. Probably weak weather artefacts or radio interference close to the noise level of the polarized intensity 
map prevented detection.

\subsection{Spectral index distribution}
\label{ss:n2976SpectralIndex}

\begin{figure}
\centering
\includegraphics[width=0.39\textwidth,clip=true]{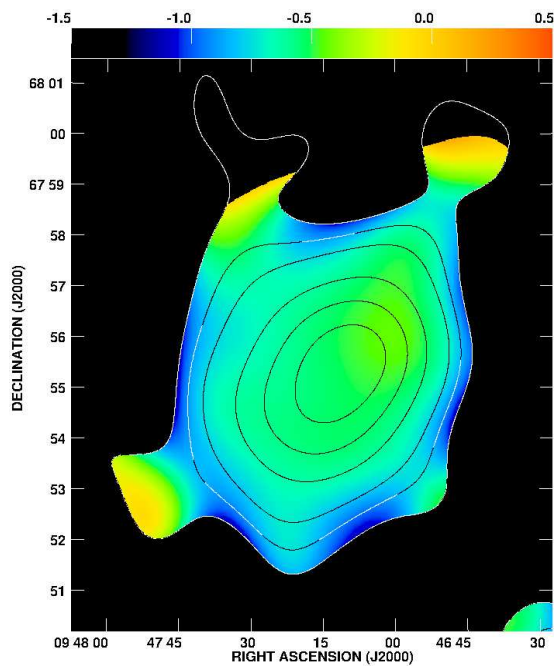}
\caption{Radio spectral index distribution between 4.85 and 1.43\,GHz in
NGC\,2976 (colours). Both maps of total intensity were convolved to a common beam of 152$"$ 
(the beam is not shown). The contours represent the total power map at 4.85\,GHz, 
their levels are: (-5, -3, 3, 5, 8, 16, 24, 32) $\times$ 450 (rms noise level) $\mu$Jy/beam.}
\label{f:n2976_SPIX}
\end{figure}

The map of spectral index distribution shown in Figure \ref{f:n2976_SPIX} was computed for NGC\,2976 using the maps at 1.43\,GHz and 4.85\,GHz, 
after convolving them to a common resolution. The input maps were cut at 3 $\times$ rms noise level. The uncertainty of the obtained values 
is less than 0.4. Despite the low resolution of the map, one of the large \ion{H}{ii} regions in the northern part of the disk, where the spectral index 
reaches a level as high as -0.4, -0.5, is visible. In the rest of the disk, the spectral index of about -0.6, -0.7 is observed. 
These values are normally encountered in the spiral arms of galaxies (see, e.g. Chy\.zy et~al. \cite{chyzy07b}). 
This could mean that the emission from the optical disk is dominated by the radiation of young relativistic electrons that are 
spiralling in a magnetic field with some small contribution from thermal emission (see Sect. \ref{ss:n2976ThermalFraction}). 
Outside the disk, the spectrum steepens sharply, reaching values of about -0.9, apparently owing to the aging of 
relativistic particles there. By way of an example, a similar effect was found for NGC\,1569 (Kepley et~al. \cite{kepley10}). 

\subsection{Distribution of rotation measure}
\label{ss:n2976RMandDP}

\begin{figure}
\centering
\includegraphics[width=0.39\textwidth,clip=true]{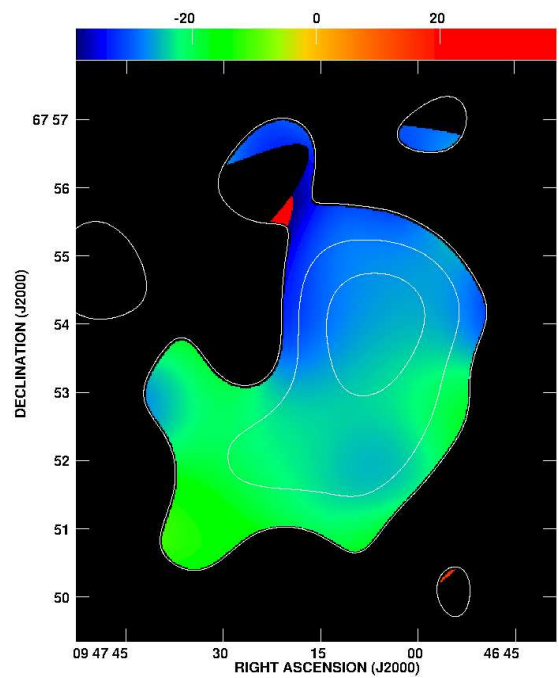}
\caption{Distribution of Faraday rotation measure (in rad m$^{-2}$, not corrected for the foreground RM) from the data at 4.85 and 1.43\,GHz in 
NGC\,2976 (colours). The contours represent the polarized intensity map at 4.85\,GHz, the levels are: (3, 5, 8) $\times$ 80 (rms noise level) $\mu$Jy/beam. 
The map resolution is 152$"$.}
\label{f:n2976_RM}
\end{figure}

In Fig. \ref{f:n2976_RM}, the distribution of rotation measure (with uncertainty less than 6 rad m$^{-2}$) is shown. 
It was calculated in a `classic' way on a pixel-by-pixel basis for the signal-to-noise ratio that was greater than 3, using the 
polarization angle maps available at two frequencies of 1.43 and 4.85\,GHz. We note from this map that the rotation measure 
observed for NGC\,2976 is mostly negative. The values of RM change smoothly from the north to the south-western direction. 
In the northern part of the object, RM has a value of about -35 rad m$^{-2}$. In moving to the south, the rotation measure 
grows from -30 to -23 rad m$^{-2}$, taking a value of about -13 rad m$^{-2}$ in the galaxy's SE part.  

When analysing the data on the rotation measure, the contribution from the Milky Way should be taken into account. One of the most standard 
ways to determine this contribution consists of using some nearby polarized background sources. In the case of  NGC\,2976,  a suitable source is 
located at  RA (J2000.0): 09$^h$48$^m$35$^s$, Dec(J2000.0): 67$^o$53$'$10$"$. Assuming that its internal RM is negligible, the foreground rotation 
measure for this source is about -35 rad m$^{-2}$. This is in  excellent agreement with the foreground rotation measure in the direction of 
NGC\,2976 that is reported by Heald et~al. (\cite{heald09}), based on their WSRT-SINGS data. This foreground RM also agrees with 
the all-sky map published by Johnston-Hollitt et~al. (\cite{johnstonHollitt04}). 

Thus, taking into account the Milky Way rotation measure, it seems    the magnetic field directed towards the observer can 
actually be seen within NGC\,2976. After correcting for the foreground RM, it ranges from zero in the north to $+22$ rad m$^{-2}$ in the 
south-east. These values are almost an order of magnitude lower than the rotation measures typically observed in normal galaxies at high frequencies 
(of about $\pm$ 100 rad m$^{-2}$, Beck \& Wielebinski \cite{beck13}) and some dwarf galaxies, e.g. NGC\,4449 (Chy\.zy et~al. \cite{chyzy00}) and NGC\,1569 (Kepley et~al. \cite{kepley10}). 
The rather low value of rotation measure found in NGC\,2976 could imply that the regular magnetic fields are weak in the object. Other possible 
explanations are that there is a low content of thermal electrons (which also affect the rotation measure) in the regions from which polarized intensity 
comes, or that the magnetic field in NGC\,2976 is aligned mainly in the sky plane.

\subsection{RM Synthesis}
\label{ss:n2976RMSynthesisResults}

\begin{figure}
\centering
\includegraphics[width=0.49\textwidth,clip=true]{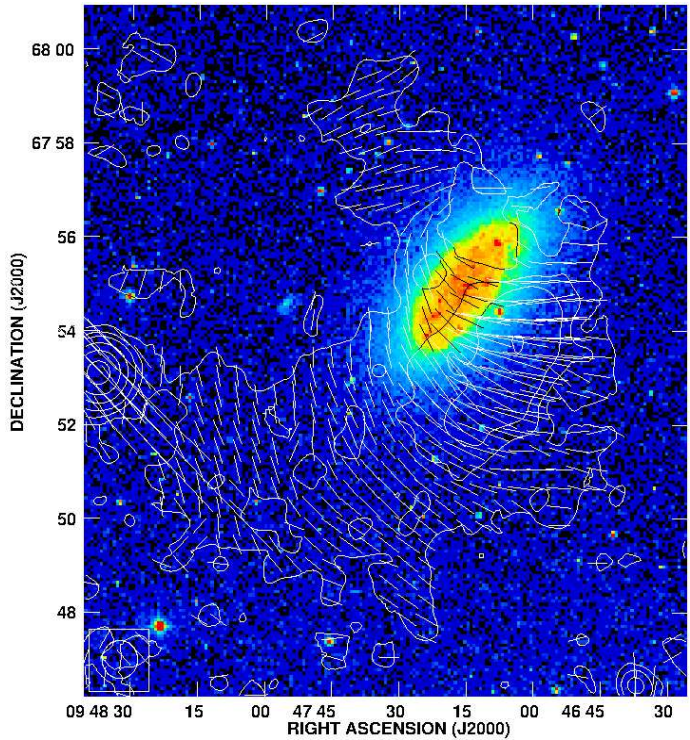}
\caption{Contours and apparent B-vectors (not corrected for Faraday rotation) of polarized intensity of NGC\,2976 from the WSRT-SINGS 
data superimposed on the DSS blue image. The contour levels are (3, 5, 8, 16, 32, 64) $\times$ 25 (rms noise level) $\mu$Jy/beam. 
The vector of 10" length corresponds to the polarized intensity of about 31.3\,$\mu$Jy/beam. The map resolution is 
48.6$"$ $\times$ 45.0$"$ HPBW.}
\label{f:n2976_wsrtPI}
\end{figure}

For NGC\,2976, spectro-polarimetric observations with the WSRT were also performed and analysed using the RM synthesis 
method (Heald et~al. \cite{heald09}) but with high spatial resolution and low sensitivity to extended radio structures. After 
re-imaging  these data (see Sect. \ref{ss:WsrtSingsReduction} for details), it was possible 
to discover polarized intensity structures (Fig. \ref{f:n2976_wsrtPI}), which had not been seen previously by Heald et~al. (\cite{heald09}). 
A comparison of the new WSRT-SINGS and VLA (see Fig. \ref{f:n2976_1_4_pi45}) polarized intensity images reveals that the distributions of 
PI in both the datasets are in very close agreement. As in the VLA, a characteristic southern bright PI 
region and an extension to the north are also visible in the WSRT image. Furthermore, these data confirm the existence of the polarized emission in the SE direction, 
which was seen after a convolution of the VLA image to the lower resolution of 45$"$.
Compared to the PI emission, the  orientations of B-vectors also agree 
well between both the observations\footnote{The central frequencies between the datasets differ slightly; for WSRT, it is 
about 1.5\,GHz, while for VLA, it is 1.43\,GHz, so the B-vectors orientations are actually slightly different.}.

\begin{figure}
\centering
\includegraphics[width=0.49\textwidth,clip=true]{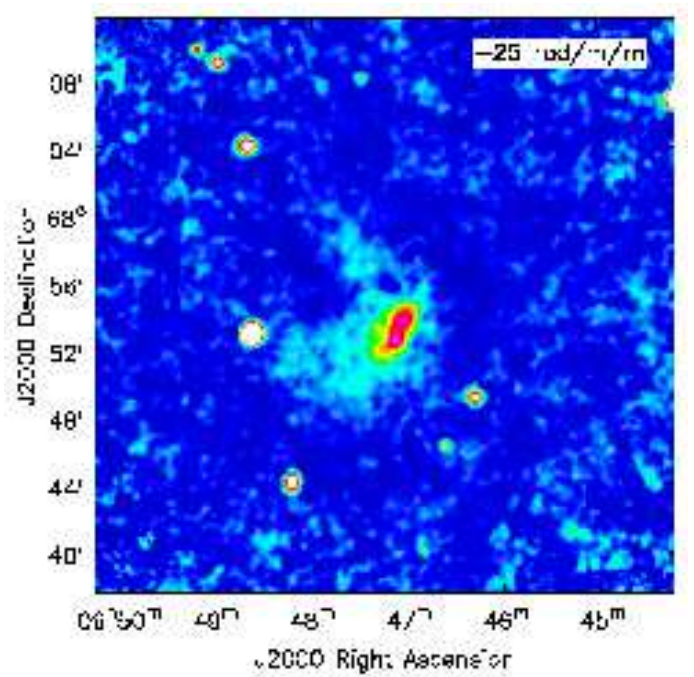}
\caption{The particular Faraday depth of -25\,rad m$^{-2}$, for which most of polarized emission 
from NGC\,2976 was detected. Note the twice as large scale of the image than that presented in Fig. \ref{f:n2976_wsrtPI}.
}
\label{f:n2976_wsrtSingleDepth}
\end{figure}

Scanning through the RM cube along the Faraday depth reveals that the polarized emission coming from NGC\,2976 is mostly 
point-like in this domain. This is in agreement with the findings of Heald et~al. (\cite{heald09}). In Fig. \ref{f:n2976_wsrtSingleDepth}, 
the slice from the cube at the depth of -25\,rad m$^{-2}$ is shown, where most of detected polarized signal can be seen.

\begin{figure}
\centering
\includegraphics[width=0.49\textwidth,clip=true]{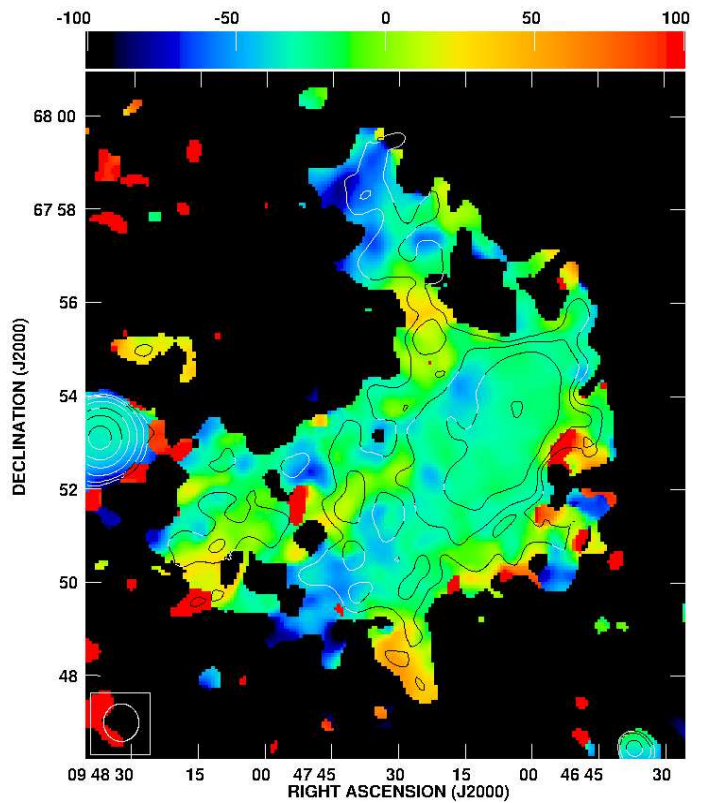}
\caption{Distribution of Faraday depths (in rad m$^{-2}$) for the maximum of polarized intensity signal for NGC\,2976 (see text for details). 
The contour levels representing the polarized intensity are: 
(4, 5, 8, 16, 32, 64) $\times$ 25 (rms noise level) $\mu$Jy/beam. The map resolution is 48.6$"$ $\times$ 45.0$"$ HPBW.}
\label{f:n2976_wsrtRM}
\end{figure}

Taking into account that there is only one point-like Faraday component in the cube, it was possible to construct a maximum Faraday depth 
map for the object (Fig. \ref{f:n2976_wsrtRM}, Sect. \ref{ss:WsrtSingsReduction}). The distribution of Faraday depths, as shown in the image ,
was cut off at the 3 $\times$ rms (of 25\,$\mu$Jy/beam) noise level of the PI map. 
We note that, according to Brentjens \& de Bruyn (\cite{brentjensAndDeBruyn05}), a total signal-to-noise ratio greater than 4 is needed to determine reliable information about Faraday 
depth using the RM Synthesis method. This limit is indicated in the image by the first PI contour. 
A comparison of the high resolution (48.6$"$ $\times$ 45$"$) maximum Faraday depth map that was obtained by applying the RM synthesis technique to the low 
resolution RM map (Fig. \ref{f:n2976_RM}) obtained by the `classical method' shows that they are similar. 
Particularly noticeable are the negative Faraday depths found in the strongly polarized SW region (this region shows a Faraday depth 
of about -25\,rad m$^{-2}$). In the northern, south-eastern, and western parts of the galaxy, where the signal-to-noise ratio is lower, the 
maximum Faraday depth distribution is more chaotic. The similarity of  both the maximum Faraday depth and RM maps indicates that Faraday 
depolarization is relatively small -- around 1\,GHz in NGC\,2976.

\section{Discussion}
\subsection{Thermal fraction}
\label{ss:n2976ThermalFraction}

\begin{figure}
\centering
\includegraphics[width=0.49\textwidth,clip=true]{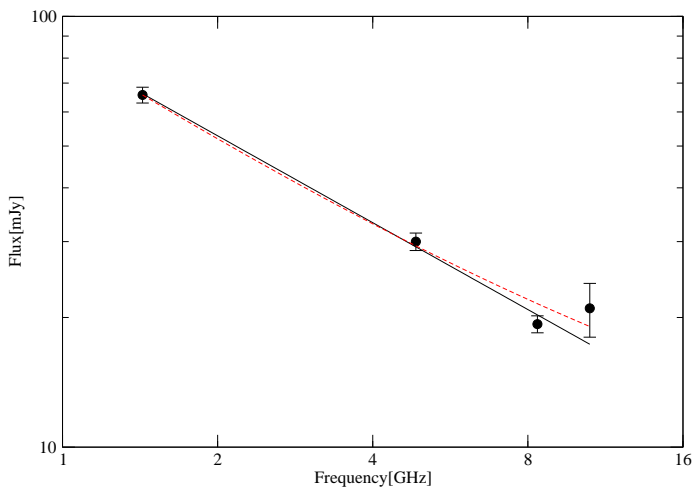}
\caption{Radio spectrum of NGC\,2976. The point at the highest frequency is from Niklas et~al. (\cite{niklas95}). 
The solid black line represents a power-law model fitted to the data. The red dashed line represents the fitted model of a thin disk in which 
f$_{th}$ = 0.16 and $\alpha_{nth}$ = -0.82 (see text for details).}
\label{f:n2976_globalSpectrum}
\end{figure}

The radio emission of normal galaxies has thermal and non-thermal components. To be able to estimate the magnetic field strength, which is 
directly related to the non-thermal radio emission, both these components must first be properly separated. The best method to  separate them 
consists of fitting a simple model of a thin optical disk to the multifrequency radio data (Niklas et~al. \cite{niklas97}). 
The model is described by the following equation:
\begin{equation}
\left(\frac{S_{\nu}}{S_{\nu_0}} \right) = f_{th}(\nu_0)\left(\frac{\nu}{\nu_0}\right)^{-0.1} + 
(1 - f_{th}(\nu_0))\left(\frac{\nu}{\nu_0} \right)^{\alpha_{nth}}
,\end{equation}
where $S_{\nu}$ and $S_{\nu_0}$ are radio fluxes for a given frequency and the frequency for which the separation is performed. Here, 
$f_{th}(\nu_0)$ is the so-called thermal fraction, being a ratio of thermal and total radio fluxes, and 
$\alpha_{nth}$ is the non-thermal spectral index. For the thermal emission, a spectral index of -0.1 is 
assumed. In the case of NGC\,2976, it was possible to use this method taking flux values from our observations at three frequencies. Moreover, as it 
was pointed out in Sect. \ref{s:n2976Results}, this galaxy was also observed in the 2.8\,cm survey of Shapley-Ames galaxies (Niklas et~al. \cite{niklas95}). 
Thus, four measurement points (Fig. \ref{f:n2976_globalSpectrum}) are available, which is a minimum that enables us to fit the above model and get the 
desired thermal fraction and non-thermal spectral index. The thermal fraction obtained in this way $f_{th}$ is 0.16 +/- 0.12 at 1.43\,GHz, 
while $\alpha_{nth}$ is -0.82 +/- 0.20. However, we note that the highest frequency data point in the spectrum (from Niklas et~al.) 
seems to be shifted upwards from the fit and should be regarded with caution.

An alternative method of separation of thermal and non-thermal emission consists of estimating the 
thermal radio flux using star formation tracers, for instance, H$\alpha$ emission (see, e.g. Niklas et~al. \cite{niklas97}; Chy\.zy et~al. \cite{chyzy07a}). 
It should be stressed, however, that H$\alpha$ is affected by internal extinction, which is usually barely known about (if at all), and which makes this method less effective. 
The extinction uncertainty is more problematic for  large well-evolved spiral galaxies than for  low-metallicity dwarfs. According to the recent 
studies, it is possible to estimate this effect by also taking into account  data on infrared emission (Calzetti et~al. \cite{calzetti07}; Kennicutt \& Evans \cite{kennicutt12}). 
By inserting the H$\alpha$ and infrared emission into the classical formula of the thermal flux estimation (Niklas et~al. \cite{niklas97}), we  obtained the 
radio thermal flux at frequency $\nu$ (see also, Jurusik et~al. \cite{jurusik14}): 

\begin{dmath}
S_{\nu}[mJy] = 2.238 \times 10^9 \left(\frac{S_{H \alpha} + 0.020 \nu_{IR} S_{IR}}{erg\ s^{-1}\ cm^{-2}} \right) \times \left(\frac{T_e}{K} \right)^{0.42} \times 
\left[ln\left(\frac{0.04995}{\nu[GHz]}\right) + 1.5ln\left(\frac{T_e}{K}\right) \right], 
\end{dmath}

\noindent where $S_{H \alpha}$ and $\nu_{IR}S_{IR}$ are H$\alpha$ and IR (multiplied by the IR frequency) fluxes, respectively. For NGC\,2976, the 
H$\alpha$ flux was taken from Kennicutt et~al. (\cite{kennicutt08}), while the IR 24\,$\mu$ flux was taken from the {\it Spitzer} satellite observations (Dale et~al. \cite{dale09}). Here,
$T_e$ was assumed to be 10$^4$\,K. The thermal fraction obtained in this way is 0.17 $\pm$ 0.02 at 1.43\,GHz, which is in  excellent agreement with the 
results of the fitting method.

This is quite unexpected, since these thermal fractions and non-thermal spectral indices are typically observed for normal spiral galaxies that are much 
more massive than NGC\,2976. Niklas et~al. (\cite{niklas97}) determined the mean $f_{th}$ of 0.08 $\pm$ 0.01 (at 1\,GHz) and an 
average $\alpha_{nth}$ of -0.83 $\pm$ 0.02 for a sample of 74 galaxies of various morphological type. 
For the dwarfs, much higher thermal fractions were found; Chy\.zy et~al. (\cite{chyzy03}) estimated that for IC\,10 $f_{th}$ reaches 
a level of 0.6 (see also, Heesen et~al. \cite{heesen11}) and for NGC\,6822, it is even as high as 0.8-0.9 at 10.45\,GHz 
(the thermal fraction of NGC\,2976 at this frequency would only be  about 0.45). Even higher $f_{th}$ were obtained for three late-type galaxies 
that are characterized by low star formation activity and explained by their weak magnetic fields and nonlinear dependency between 
nonthermal emission and SFR (Chy\.zy et~al. \cite{chyzy07a}). 

In the same way as the thermal fraction, the global spectral index of NGC\,2976 (-0.67 $\pm$ 0.03) also follows the value of -0.74 $\pm$ 0.12 
found for a sample of normal galaxies (Gioia et~al. \cite{gioia82}). 

\subsection{Magnetic field strength}
\label{ss:n2976BfieldStrength}

After separating the thermal and non-thermal emissions, it was possible to estimate the strength of the magnetic field in the investigated object. 
The formulas given by Beck \& Krause (\cite{beck05}) were used for this purpose, with the assumption of an energy-density equipartition between the total field and cosmic 
rays (CR). A typical disk thickness (corrected for the projection effects) of 1\,kpc and the proton-to-electron ratio of 100 were assumed in the 
calculations (using the non-thermal spectral index of 0.83 from Niklas et~al. \cite{niklas97}). The mean non-thermal surface brightness was measured in the 
area that was determined by the requirement of radio emission being at a level of at least 3$\sigma$ noise level at 1.43\,GHz (see Sect. \ref{s:n2976Results}), 
giving the total magnetic field strength B$_{tot}$ of 6.6 $\pm$ 1.8\,$\mu$G. 

The total magnetic field strength obtained is higher than the mean value of $<$ 4.2 $\pm$ 1.8\,$\mu$G found for a sample of  Local Group dwarfs 
(Chy\.zy et~al. \cite{chyzy11}), but lower than the average value of 9 $\pm$ 2\,$\mu$G that was estimated for a sample of 74 galaxies of various morphological types 
(Niklas \cite{niklas95PhD}). However, we note that the Niklas sample also included  interacting objects, which are usually characterized by 
higher magnetic fields strengths than those in normal galaxies (Drzazga et~al. \cite{drzazga11}). In spiral galaxies with moderate star formation activity, 
e.g. in M\,31 and M\,33, and so more similar in this sense to NGC\,2976 (for which SFR $\approx$ 0.12 M$\odot$/yr, see Table \ref{t:wsrtSingsIntroTab}), the field strengths of only 
about 6\,$\mu$G were found (Beck \& Wielebinski \cite{beck13}).  

Another important parameter to describe galactic magnetic field properties is the degree of field order, defined as a ratio of the ordered (B$_{ord}$) and 
random (B$_{ran}$) magnetic field components. This is a useful measure of the net production of the ordered field. The B$_{ord}$ estimated 
from the data about polarized flux (using formulas of Beck \& Krause \cite{beck05}) for NGC\,2976 is 2.8 $\pm$ 0.8\,$\mu$G. This is a typical value for the spirals, in which ordered fields in the 
range of 1 -- 5\,$\mu$G are observed (Beck \& Wielebinski \cite{beck13}). Along with the random field component 
(which for NGC\,2976 is B$_{ran} = $ 6.0 $\pm$ 1.6\,$\mu$G), this gives a very high value of degree of field order of 0.46 $\pm$ 0.17. 
Consequently, with respect to the strength of magnetic field components, NGC\,2976 resembles well-defined spiral galaxies rather than dwarfs.

\subsection{Generation of magnetic field}
\label{ss:n2976Dynamo}

As was shown in Section \ref{s:n2976Results}, non-zero rotation measures were found for NGC\,2976, which possibly suggest that this object hosts a 
large-scale partially regular magnetic field. One of the mechanisms to explain the generation of this field is a large-scale magnetohydrodynamic dynamo 
(e.g. Widrow \cite{widrow02}). To investigate if this process is at work in this galaxy, we can estimate its efficiency, 
usually described by a dynamo number (Shukurov \cite{shukurov07})
\begin{equation}
D \approx 9\frac{h_{0}^{2}}{u_{0}^{2}} s \omega \frac{\partial \omega}{\partial s}
\label{e:dynamoEq}
,\end{equation}
where, $h_{0}$ is the vertical-scale height of the galactic disk, $u_{0}$ is the velocity of turbulent motions, $s$ is the 
radial distance from the centre of the galaxy, $\omega$ is the angular velocity and $s \frac{\partial \omega}{\partial s}$ 
is the shear rate. To estimate this number for NGC\,2976, we assumed a scale height of $0.5$\,kpc, typical for dwarf 
galaxies (e.g. Chy\.zy et~al. \cite{chyzy11}) and $u_{0}$ from \ion{H}{i} observations of 11\,km s$^{-1}$ (Stil \& Israel \cite{stil02b}). 
The dynamo number was computed for a radius of 2.0\,kpc, within which the rotation curve can be approximated as flat (as it was suggested 
by Stil \& Israel \cite{stil02b}, but questioned by de Blok et~al. \cite{deBlok08}). At this distance, the rotation velocity reaches a 
value of about 71\,km s$^{-1}$ (Stil \& Israel \cite{stil02b}). Although it rotates with a velocity that is somewhat slower than that of typical 
spirals it is still faster than the low-mass galaxies, NGC\,4449 and the SMC (50 and 60\,km s$^{-1}$, respectively). As the rotation of the disk 
in NGC\,2976 is also more regular than in  SMC and NGC\,4449, it can be expected that the large-scale dynamo process can operate in 
this object.

Indeed, an estimation for NGC\,2976 of a shear rate of about 36\,km s$^{-1}$ kpc$^{-1}$ and the 
absolute value of the dynamo number of about 23 shows that this is comparable to the dynamo number that was obtained for the Milky Way (of about 20) at 
the distance of the Sun (Shukurov \cite{shukurov07}). It exceeds the critical limit $|D_{critical}| \sim  8-10$, below which the 
large-scale dynamo is inefficient (Shukurov \cite{shukurov07}). This value does not seem to be typical for dwarf galaxies, for which 
mostly subcritical dynamo numbers have been obtained (Chy\.zy et~al. \cite{chyzy11}; Mao et~al. \cite{mao08}). 
Even for the Large Magellanic Cloud, which is much larger and more massive than NGC\,2976, Mao et~al. (\cite{mao12}) 
estimated $|D|$ as $\sim$10 only. Thus, with the linear size of about 6\,kpc and \ion{H}{i} mass 
of 1.5x10$^8$\,M$_\odot$ (Stil \& Israel \cite{stil02a}, Table \ref{t:wsrtSingsIntroTab}), NGC\,2976 is several times less massive 
than SMC and NGC\,4449 and could set the lowest mass threshold for a large-scale dynamo to work. 

Based on simple estimates for the turbulence transport coefficients, it has often been argued for a too long  growth time of the $\alpha-\omega$ dynamo 
(see e.g. Shukurov \cite{shukurov07}). But direct simulations of the turbulent ISM by Gressel et~al. (\cite{gressel08}) led to 
growth times of approximately 200\,Myr for an $\alpha-\omega$ dynamo process. A similar growth time is found 
by Hanasz et~al. (\cite{hanasz04}) with cosmic ray-driven steering of the ISM. In this type of dynamo process, supernova explosions 
produce cosmic rays, the pressure from which leads to the increased efficiency of the $\alpha$ effect, which is responsible for generating the poloidal component 
of magnetic field. It has recently been successfully applied  in simulations of magnetic fields in spiral (Hanasz et~al. \cite{hanasz09}) and 
barred galaxies (Kulpa-Dybe\l~et~al. \cite{kulpa11}), yielding dynamo growth timescales of 270\,Myr and 300\,Myr, respectively. The CR dynamo is also considered to be a particularly attractive mechanism  for 
dwarfs, and is able to account for their regular magnetic 
fields (Gaensler et~al. \cite{gaensler05}; Kepley et~al. \cite{kepley10}, Mao et~al. \cite{mao12}). This explanation is 
further supported by  recent global numerical simulations of dwarfs performed by Siejkowski (\cite{siejkowski12}), who find that the magnetic 
field can be amplified in low-mass objects only if their rotational velocity exceeds 40\,km s$^{-1}$, However, a velocity of 
at least 60\,km s$^{-1}$ is needed for the efficient generation of a magnetic field, and for such fast rotators, their dynamo action depends 
mainly on rotational velocity. 

By confronting the maximum rotational velocity of NGC\,2976 with these results, we can see that the cosmic-ray driven dynamo is actually able to 
operate efficiently in this object with e-folding amplification time not greater than 400\,Myr (cf. Siejkowski's models 
with v$_{rot} \ge 60$\,km s$^{-1}$). Furthermore, the strength of the magnetic field in NGC\,2976 (Sect. \ref{ss:n2976BfieldStrength}) is roughly in  
agreement with the simulations, where for the fast rotating models this is at the level of a few $\mu$G. 

\subsection{Magnetic field morphology of NGC\,2976}
\label{ss:N2976FieldStructure}

\begin{figure}
\centering
\includegraphics[width=0.49\textwidth,clip=true]{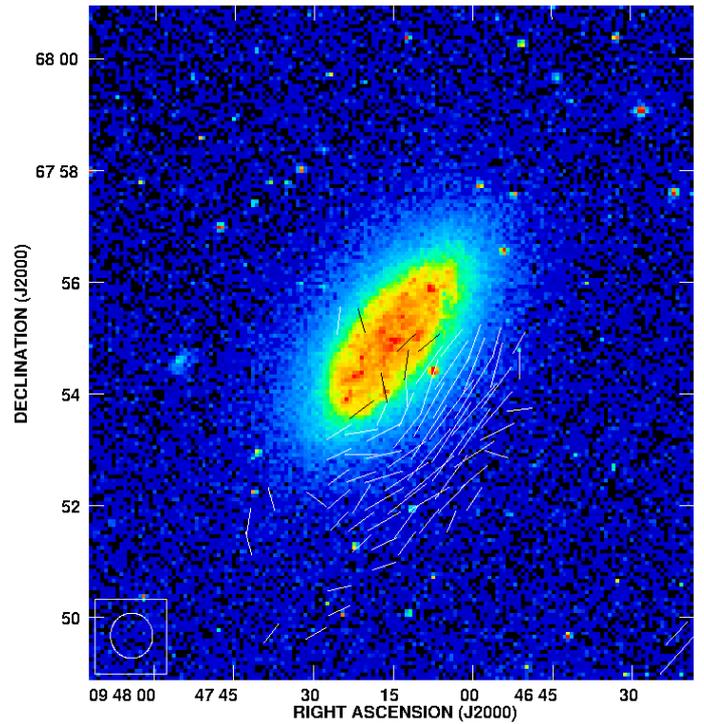}
\caption{Orientations of intrinsic magnetic field vectors (plotted for a signal-to-noise ratio of at least 5 in PI) for NGC\,2976 
from the reprocessed WSRT-SINGS data superimposed on the DSS blue image. 
The vector of 10$"$ length corresponds to the polarized intensity of 50\,$\mu$Jy/beam. The beam size is 
48.6$"$ $\times$ 45.0$"$ HPBW.}
\label{f:N2976AndSingsBfields}
\end{figure}

Apart from the field strength,  its morphology is also crucial when discussing the large-scale dynamo process. 
In Fig. \ref{f:N2976AndSingsBfields}, a distribution of magnetic field orientations (B-vectors corrected for Faraday rotation) is 
presented for NGC\,2976. The map was computed only for the regions where signal-to-noise ratio was high enough 
(see Brentjens \& de Bruyn \cite{brentjensAndDeBruyn05}). The most surprising are the almost azimuthal (pitch angle around zero degrees) 
directions of the field in the southern part of the object, where the maximum of polarized intensity is observed. 
Hence, we  suspect that the magnetic field in NGC\,2976 could not have been generated by the galactic-scale dynamo 
action alone, since  a spiral morphology of magnetic fields would have arisen as a result. So far, this kind of field has been observed most clearly in the 
ring galaxy NGC\,4736 (Chy\.zy \& Buta \cite{chyzyButa08}), which does not exhibit any strong density waves,  like NGC\,2976. 
The nearly azimuthal magnetic field in the south, along with the enhanced polarized intensity and no corresponding enhancement of total power 
emission (cf. Fig. \ref{f:n2976_1_4_tpHI} and Fig. \ref{f:n2976_1_4_piHalpha}) imply a field compression in this part of the galaxy. 
The degree of field order of 114 $\pm$ 48\,\% that is estimated for this area is very high, since it is 2.5 times higher than the 
average value (Sect. \ref{ss:n2976BfieldStrength}).

\subsection{Tidal interactions or ram pressure?}
\label{ss:n2976TidalvsRamPressure}

Ram pressure caused by the intra-group medium of the M\,81 galaxy group is one possibility of accounting for the magnetic field 
structure of NGC\,2976. This mechanism is most often observed in clusters of galaxies, e.g. in the Virgo cluster 
(see, e.g. Vollmer et~al. \cite{vollmer13}). In  galaxy groups, the relative speed of  
objects with respect to the surrounding medium and its density (the quantities that the ram pressure depends on) are usually one order 
of magnitude lower than for the clusters. Recently, Bureau \& Carignan (\cite{bureau02}) have considered the influence of ram pressure on the 
galaxy HoII, which is located about 0.5\,Mpc from the core of the M\,81 group. They argue that, for this low-mass object, the density of the ambient 
medium of the M\,81 group and its relative velocity are high enough to cause its gas stripping. Thus, Bureau \& Carignan's argument 
is all the more valid for NGC\,2976, the distance of which to the group centre is lower (and was supposedly even more so, in the past). 
The ram pressure affects only the gaseous component (hence the magnetic field frozen into the plasma), leaving the stellar 
disk untouched. In the polarized emission,  a compressed field is usually observed at the windward side, forming 
a polarized ridge and a diffuse extension in the opposite side of galactic disk (see, e.g. Vollmer et~al. \cite{vollmer10}, 
Vollmer et~al. \cite{vollmer13}). This kind of effect could explain the case of 
NGC\,2976 since, in fact, the structure of its magnetic field closely resembles  those observed in some  Virgo cluster spirals, particularly 
in NGC\,4501 (Vollmer et~al. \cite{vollmer08}), which has a similar inclination to the galaxy in question. 

However, for the ram pressure, a more narrow shock front is to be expected and not a wide one, as observed in the investigated object. 
Compared to NGC\,4501, the polarized emission in NGC\,2976 is extended farther beyond the optical disc. 
Also, no  index  flattening is seen in NGC\,2976 (Sect. \ref{ss:n2976SpectralIndex}) in the region of maximum polarized emission, which is
expected for the strong ram pressure (see Vollmer et~al. \cite{vollmer13}). 
Moreover, recent studies of nearby low-mass galaxies show that NGC\,2976 fits into the general 
trend of radio-infrared correlation that is determined for more massive spirals (Jurusik et~al. \cite{jurusik14}). For Virgo cluster galaxies that
exhibit ram pressure stripping, deviations from the correlation were observed 
(Murphy et~al. \cite{murphy09}; Vollmer et~al. \cite{vollmer13}), which can be accounted for by an 
enhanced radio emission owing to gas compression. Thus, again, in the case of NGC\,2976, the ram pressure effects, if any, must be 
weak.
Also the morphology of neutral hydrogen in the southern part of NGC\,2976, while being coincident with the PI maximum, it is not so clearly 
defined as in the other galaxies that are known to have been stripped, e.g. in the Virgo spirals, NGC\,4501 (Vollmer et~al. \cite{vollmer08}) and 
NGC\,4654 (Soida et~al. \cite{soida06}). Even the gas compression in HoII (Bureau \& Carignan \cite{bureau02}) seems to be more pronounced 
than in NGC\,2976. Taking the present location of NGC\,2976  into account with respect to the group core, it is quite plausible that this 
object has already passed the peak of ram pressure. In fact, Otmianowska-Mazur \& Vollmer (\cite{otmianowska03}) define two phases that are related to  
ram pressure, i.e. a compression phase and a phase of re-accretion, when gas falls back onto the galactic disk in a spiral-like manner. 
In their simulations, in the early stages of gas re-accretion the polarized intensity is also enhanced. Moreover, maxima of the gas 
distribution outside of the disk should occur in this phase, which is actually observed in the case of NGC\,2976.
Another possibility, related to the ram pressure, is the influence of draping flows which modify the large-scale dynamo action and lead to 
an asymmetric PI distribution (Moss et~al. \cite{moss12}). However, in this case, relatively strong ram pressure is also needed.

\begin{figure}
\centering
\includegraphics[width=0.49\textwidth,clip=true]{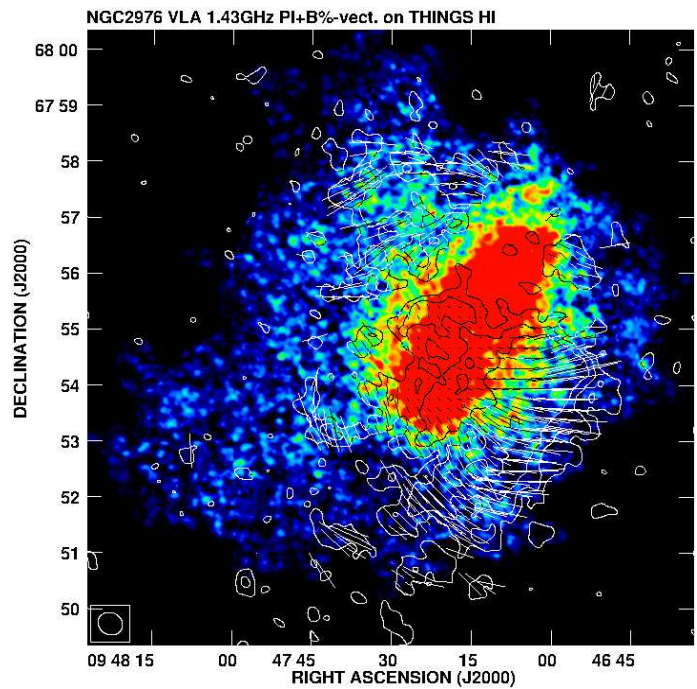}
\caption{Contours of polarized intensity and apparent B-vectors (not corrected for Faraday rotation) of polarization degree of NGC\,2976 at 1.43\,GHz 
(obtained with Robust=1 weighting) superimposed on the \ion{H}{i} map (from the THINGS survey, Walter et~al. \cite{walter08}). The contour levels 
are (3, 5, 8) $\times$ 9 (rms noise level) $\mu$Jy/beam. The vector of 10" length corresponds to the polarization degree of 12.5\%. 
The polarized intensity map resolution is 26.8$"$ $\times$ 23.5$"$ HPBW (the beam position angle is 62 degrees). 
The \ion{H}{i} map resolution is 7.4$"$ $\times$ 6.4$"$ HPBW (the beam position angle is 72 degrees).}
\label{f:n2976_piOnHI}
\end{figure}

These potential difficulties with the ram pressure scenario suggest that, in NGC\,2976, tidal interaction could play a major role. 
It is thought that the galaxy could have been influenced by M\,81 in the past, during its passage close to the group centre. 
At present, NGC\,2976 is located 
south-west of the M\,81 group's core. This kind of interaction, between M\,81 and NGC\,2976, could have occurred more than 1.3\,Gyr ago, as  
has been  approximately estimated  by Williams et~al. (\cite{williams10}). In the high-sensitivity maps of 
distribution of neutral hydrogen, a long strip of \ion{H}{i} that connects the object with the group centre, was detected 
(Chynoweth et~al. \cite{chynoweth08}). Hence, some tidal interactions actually could have taken 
place (Appleton et~al. \cite{appleton81}, Appleton \& van der Hulst \cite{appleton88}, Chynoweth et~al. \cite{chynoweth08}). 
The tidal origin of the magnetic field morphology, which is observed in NGC\,2976, explains its asymmetric PI distribution, seen at 1.43\,GHz 
and 4.85\,GHz and the location of the polarized emission beyond the optical disk and well within the \ion{H}{i} distribution 
(Fig. \ref{f:n2976_piOnHI}). The galaxy also seems not to  have a strong ordered magnetic field in its disk. 
Altogether, these observables have been recognized by Vollmer et~al. (\cite{vollmer13}) as signatures of tidal interactions 
in the Virgo cluster spirals.

Usually, the degree of field order is regarded as an indicator of tidal interactions, since it is statistically lower for 
the interacting objects than for the non-interacting ones (Drzazga et~al. \cite{drzazga11}). 
This is not the case for NGC\,2976, for which the degree of field order mentioned above 
(Sect. \ref{ss:n2976BfieldStrength}) has a value that is typically found  in  non-interacting galaxies. 
This is not too surprising, since    obviously interacting galaxies are known about, e.g. NGC\,4254, for 
which the degree of field order is similar to those observed in normal objects (Chy\.zy \cite{chyzy08}).

However, there are also some problems with attributing the unusual structure of the magnetic field of 
NGC\,2976 to  tidal forces. One of them is its ideally symmetric stellar disk and halo, which 
can be an apparent indication that the galaxy has not been disturbed for a long time (Bronkalla et~al. \cite{bronkalla92}). 
This makes the tidal interactions scenario more doubtful, but not impossible, since magnetic structures can involve quite a long `memory' of  
past events (from even  1\,Gyr, see Soida et~al. \cite{soida02} and ref. therein). 
This memory is longer when compared to the other components of the ISM and the stars. The best example of 
a mismatch between the structure of the stellar disk and the structure of the field was so far found  in the 
flocculent galaxy, NGC\,4414 (Soida et~al. \cite{soida02}). In this high-mass and non-interacting object, 
as in NGC\,2976, a symmetric flocculent disk can be seen, suggesting that it has not been strongly affected recently. 
However, the magnetic field contradicts this altogether, since it 
shows a complicated morphology as well as a strong RM jump in the southern part of the object, which is not expected  in a non-disturbed galaxies (Soida et~al. \cite{soida02}). Indeed, recent studies of de Blok et~al. 
(\cite{deBlok14}) indicate that NGC\,4414 could have taken part in a minor interaction with low mass galaxy in 
the past.

From the discussion presented above, we can propose the following scenario, which reasonably explains the structure of the magnetic field in 
NGC\,2976. More than 1\,Gyr ago, the galaxy took part in  gravitational interactions between galaxies in 
the group centre. At this time, tidally induced radial inflow of gas occurred, leading to the outside-in truncation of 
star formation, specifically for NGC\,2976 (Williams et~al. \cite{williams10}). As shearing motions and compression are both associated with 
gas inflow, hence like ram pressure, they are able to produce an ordered or anisotropic magnetic field from its random component, 
while not disturbing the stellar disk. These external influences were probably relatively weak, and thus could have had more impact on 
the least dense ISM phases. The currently observed anomalies in the magnetic field configuration of the object would be a result of the `memory' 
of these past interaction events.

Williams et~al. (\cite{williams10}) also consider  that the ram pressure effect causing the stripping of halo gas as an alternative 
(or additional process), which is  able to explain the star formation history of NGC\,2976. From our studies presented above, it seems that this 
scenario is less likely than the gas redistribution owing to tidal interactions. However, we cannot exclude that  ram pressure in its 
late phase of \ion{H}{i} gas and magnetized plasma re-accretion could also have played some part in shaping the magnetic field in this object.

Detailed 3D numerical MHD simulations are required to evaluate this scenario. NGC\,2976 is a quite suitable 
object for this type of simulation, since it does not exhibit any strong density waves to influence the structure of its magnetic field. Moreover, from the 
observational point of view, sensitive X-ray data are needed to detect the hot gas halo around this object. This kind of halo, if any, could have a 
significant impact on the effectiveness of the ram pressure process in NGC\,2976.

\subsection{Pure disk dwarf galaxies as a potential sources of  IGM magnetic fields}
\label{ss:NGC2976IGM}
The magnetization of intergalactic space is still an open problem in modern astrophysics (see Sect. \ref{s:n2976Introduction}). As it was 
shown in Sect. \ref{s:n2976Results}, NGC\,2976 exhibits a strong TP radio emission, extending up to 4.5\,kpc (without correction for the 
inclination angle), thus in a general agreement with Chy\.zy et~al. (\cite{chyzy11}) predictions that are based on the assumption that the star formation was continuous in galaxies, 
which is the case for NGC\,2976 (Williams et~al. \cite{williams10}). While it is not sufficient to regard NGC\,2976 as an effective provider of magnetic field 
into the IGM, it is worthwhile  noting that this galaxy shows also an extended polarized emission of a similar extension as the total power one 
(5\,kpc, if measured in the higher sensitivity map, Fig. \ref{f:n2976_1_4_pi45}). Thus, apart from the random component of the magnetic field, NGC\,2976 
can also supply  well-ordered fields to the IGM, at least within its closest neighbourhood. Moreover, the calculations presented above show that this object is 
not only able to host a large-scale dynamo, but also creates favourable conditions for it to act at a level that is typically found in spiral galaxies. 
This, along with its low mass, makes NGC\,2976 one of the best known instances of providers of a regular (not just ordered) intergalactic magnetic field. 
Even in the galaxy NGC\,1569, of \ion{an H}{i} mass similar to NGC\,2976, with a huge radio halo that extends up to 4\,kpc (as estimated in Fig. 1 from 
Kepley et~al. \cite{kepley10}), no highly polarized signal was detected at 1.4\,GHz (at higher frequencies, a stronger PI is visible, but it is less extended). 
Therefore, this object is as efficient as NGC\,2976 in providing random fields to its neighbourhood, but it is much less efficient in providing 
ordered and regular fields.

Hence, to efficiently magnetize the whole M\,81 group with a radius of about 211\,kpc (Karachentsev \cite{karachentsev05}), or at least its central 
part, more objects from the pure disk dwarf class are needed. However, finding objects like NGC\,2976 is difficult because its morphology seems to 
be unique. It does not look like any other known nearby galaxy, featuring a bright sharply-edged disk, which is perhaps 
just a short-lived stage within its entire evolution (Williams et~al. \cite{williams10}). In fact, within the M\,81 group, which consists of 29 galaxies 
(Karachentsev \cite{karachentsev05}; or 36 according to Karachentsev \& Kaisin \cite{karachentsev07}), only one galaxy  can be found that is similar to NGC\,2976, i.e. NGC\,4605 
(Karachentsev \& Kaisin \cite{karachentsev07}). This galaxy also has a well-defined stellar disk and is only slightly larger (with its disk of about 8\,kpc) than 
NGC\,2976\footnote{It should be noted that NGC\,4605 has not been classified as the M\,81 group member in the Karachentsev (\cite{karachentsev05}) work. However, in a 
later paper Karachentsev \& Kaisin (\cite{karachentsev07}), it was included in the sample of  group galaxies studied in H$\alpha$.}. In short, the small number of the 
galaxies of this type makes their significant role for the group's magnetization process rather doubtful. It should be emphasised here that this 
conclusion is  based only on the observed 1.43\,GHz radio envelope. If we assume that the magnetic field of NGC\,2976 extends to at least a distance 
where its \ion{H}{i} emission is visible (Chynoweth et~al. \cite{chynoweth08}), the volume which the object could effectively magnetize would be much larger. 
To detect the full envelope of synchrotron emission that is related to this type of field, a new generation of radio telescopes are needed. LOFAR is one 
of the best-suited instruments for this task since it can observe at low frequencies within the range of 30 -- 240\,MHz. Thus, it would be able to 
detect a weak magnetic field that is highlighted by low-energy electrons leaving the disk of NGC\,2976.

\section{Summary and conclusions}
For NGC\,2976, a pure-disk dwarf galaxy, deep polarimetric observations at 1.43\,GHz, 4.85 and 8.35\,GHz were performed and then supplemented with 
re-imaged WSRT-SINGS data. The radio images that were obtained were compared with data in other spectral domains, which led to the following conclusions:

\begin{itemize}

\item A large total-power and polarized radio envelope surrounding the object was discovered. 

\item The thermal fraction that was estimated using two different methods (giving the complementary results) is about 0.17 (at 1.43\,GHz), 
which is similar to that found in typical spiral galaxies. Also the total magnetic field strength (of about 7\,$\mu$G), its 
ordered component (of about 3\,$\mu$G), and degree of field order (of about 0.46), are typical for spiral galaxies, despite 
the low mass of NGC\,2976.

\item The analysis of rotation-measure distribution suggests the possible existence of a coherent, large-scale dynamo-generated 
magnetic field in NGC\,2976. 
The dynamo number calculated for this low-mass galaxy greatly exceeds the critical value, indicating that, in this object, 
a large-scale $\alpha$ -- $\omega$ dynamo can indeed work efficiently. 
This is quite a surprising result, since sub-critical dynamo numbers were usually obtained for  other dwarf galaxies  studied to date.

\item Re-imaging of the WSRT-SINGS data for NGC\,2976 reveals a much more extended polarized signal than in the Heald et~al. (\cite{heald09}) work, 
which is compatible with our sensitive VLA 1.43\,GHz observations. The  RM synthesis shows a single point component in the Faraday 
depth space. This confirms an RM distribution that was obtained by  classical radio polarimetry from  two frequencies (1.43\,GHz and 4.85\,GHz).

\item The morphology of the magnetic field found in NGC\,2976 does not resemble that expected for  pure dynamo action since it features a 
polarized ridge in the south. This, along with an undisturbed optical appearance 
of the object, suggests that its magnetic field may `remember'  past tidal interactions with the M\,81 galaxy (or the centre group galaxies). 
It is also possible that the ram pressure of the IGM could have further modified the morphology of the magnetic field observed in NGC\,2976.

\item The number of galaxies with a morphology similar to NGC\,2976 in the M\,81 group is not big enough to consider these objects an 
efficient source of magnetization of the intra-group medium. These galaxies (taking into account only the extension of their radio emission) 
can provide an ordered (or even regular) magnetic field to their neighbourhood up to distance of 5\,kpc.

\item A new weighting scheme, based on the quality of the individual frequency channels, for the RM synthesis algorithm was proposed and 
investigated. The application of this new weighting to the simulated data, as well as to the observed data, results 
in an improvement of the signal-to-noise ratio in the Faraday depth space.

\end{itemize}

\begin{acknowledgements}
We thank an anonymous referee for helpful comments and suggestions.
This research has been supported by the scientific grant from the Polish National Science Centre (NCN), decision no. DEC-2012/07/N/ST9/04146.
JSG's research on magnetic fields is partially supported through a Rupple Bascom Professorship that is generously funded by the University of
Wisconsin-Madison. We acknowledge the use of the HyperLeda (http://leda.univ-lyon1.fr) and NED (http://nedwww.ipac.caltech.edu) databases.
\end{acknowledgements}

\appendix

\section{Application of radio interferometry weighting to RM synthesis}
\label{ch:newWeighting}

Processing spectro-polarimetric data using RM synthesis involves the application of frequency channel weighting.
The method currently used (the so-called classic weighting scheme) consists of assigning weights of 1 to all 
the observed frequency channels and 0 to the others.

\begin{figure*}[t]
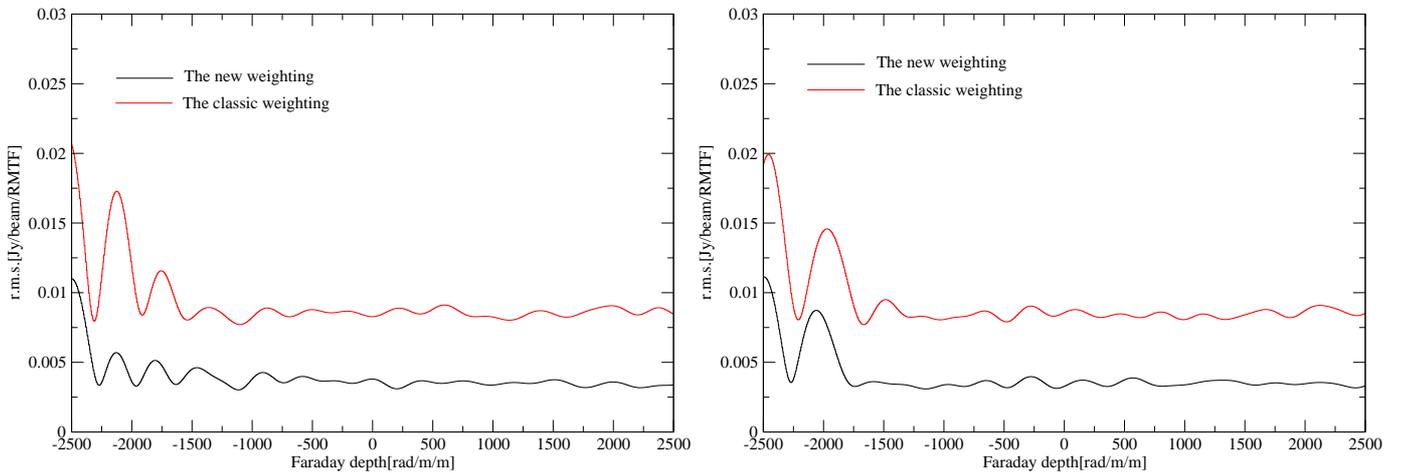

\centering
\includegraphics[angle=0,width=0.49\textwidth,clip=true]{RMrmsQNaturalClassical.eps}
\includegraphics[angle=0,width=0.49\textwidth,clip=true]{RMrmsUNaturalClassical.eps}
\caption{Rms noise levels for real (left) and imaginary (right) parts of the Faraday dispersion function. The measurements were 
performed in the same region as for the maps in the frequency domain.}
\label{f:RMQUrmsSimulations}
\end{figure*}

Obtaining the highest possible resolution in the Faraday domain requires observations in a wide spectral band. 
An example of this can be found in the WSRT-SINGS survey (Heald et~al. \cite{heald09}), where data were 
collected in the 18 and 22\,cm bands, simultaneously. As only the weights of 1.0 or 0.0 are assigned to the particular 
frequency channels according to the classic weighting, it is highly desirable to obtain a uniform rms sensitivity 
across the observed band. However, in practice, it can be difficult to meet this criterion, which was the case also 
for the survey. Furthermore, sensitivities of individual channels 
within a single frequency band can considerably differ. This is often due to problems that are related to RFI or calibration errors. 
Thus, treating all (usable) frequency channels in the same way by assigning them the same weights of 1.0 does not seem  to be 
the most adequate solution. In this appendix, we present a novel approach to weighting data in rotation 
measure synthesis, which permits the individual channel sensitivities to be taken into account.

\subsection{Introduction to the new weighting scheme}
Rotation measure synthesis is similar to radio interferometry (Brentjens \& de Bruyn \cite{brentjensAndDeBruyn05}). 
As such, one can try to write a formula for the RM Synthesis weight of j-th frequency channel in a similar way to the single
visibility data point in imaging aperture synthesis observations (e.g. Briggs et~al. \cite{briggs99}):

\begin{eqnarray}
w_j(\lambda^2_j) = \frac{R_j(\lambda^2_j)T_j(\lambda^2_j)D_j(\lambda^2_j)}{max(R_j(\lambda^2_j)T_j(\lambda^2_j)D_j(\lambda^2_j))}
\label{e:newWeightingFormula}
.\end{eqnarray}
In radio interferometry these three elements (of imaging weight) are defined as follows: $R_j$ describes reliability of a given 
visibility (it is related to the data quality), $T_j$ is a tapering function (related to the position of a given data point in the 
uv-plane), and $D_j$ is a density-weighting function. The last coefficient can be used to compensate a non-uniform distribution of data 
points in the uv-plane (for more details, see Briggs et~al. \cite{briggs99}). 

In RM synthesis, we can apply a similar weighting strategy and assume that $R_j$ will describe a quality of an individual data point 
(i.e. Q and U Stokes parameters for a given frequency channel). We propose to approximate $R_j$ by the following formula: 

\begin{eqnarray}
R_j(\lambda^2_j) = \frac{1}{\sigma^2_Q(\lambda^2_j) + \sigma^2_U(\lambda^2_j)} 
\label{e:RjFormula}
,\end{eqnarray}
where, $\sigma_Q(\lambda^2_j)$ and $\sigma_U(\lambda^2_j)$ are the rms noise levels measured for a particular Q and U frequency channel, 
respectively. We  chose $1/\sigma^2$ weighting, as it is widely used in statistical analysis and radio-interferometry 
(e.g. Briggs \cite{briggs95}, Briggs et~al. \cite{briggs99}). Substituting $T_j$ = $D_j$ = 1 in formula \ref{e:newWeightingFormula} 
gives an equivalent of the so-called natural weighting of uv data points in aperture synthesis\footnote{We do not consider here  cases where $T_j \neq$ 1 and $D_j \neq$ 1}. 
To test the new concept, we performed 
simple simulations of observations which are described below.

\subsection{Simulations}

\begin{figure}
\centering
\includegraphics[clip, width=0.5\textwidth,clip=true]{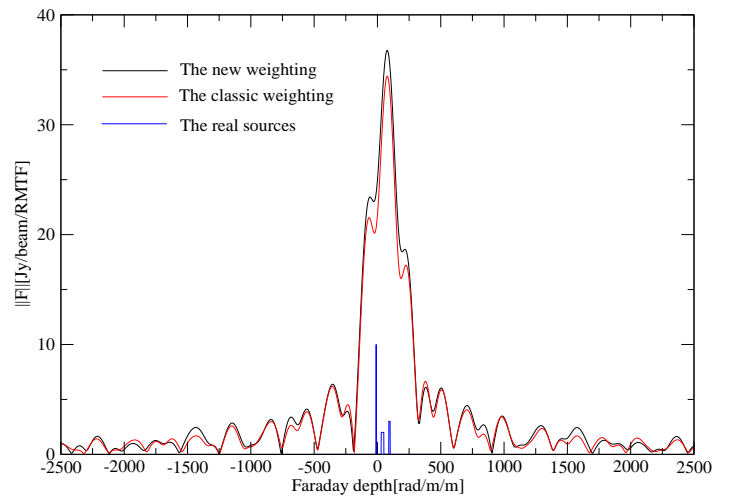}
\caption[Faraday spectra obtained with the new and the classic weighting.]
{Faraday spectra obtained with the new (black) and the classic weighting (red) for the given sources (blue).}
\label{f:FaradayDispersionSimulations}
\end{figure}

In the simulations, we modelled the following three sources in the Faraday depth 
domain\footnote{Properties of the sources are exactly the same as in the Appendix B of Brentjens \& de Bruyn (\cite{brentjensAndDeBruyn05}).} 
(see Fig. \ref{f:FaradayDispersionSimulations}): a pointlike source located at a Faraday depth of -10\,rad m$^{-2}$ with 
flux of 10\,Jy, a source extending from +30 to +50\,rad m$^{-2}$ with total flux of 40\,Jy, and a source extending from 
+90 to +100\,rad m$^{-2}$ with a total flux of 30\,Jy. These three sources (in the Faraday depth domain) are each seen as 
a single pointlike source in the sky plane.

For these sources, we simulated the spectro-polarimetric observations with a frequency setup similar to that used in 
the WSRT-SINGS survey (see Heald et~al. \cite{heald09}). Thus, we obtained two frequency bands (18\,cm and 22\,cm), each with 512 frequency 
channels. To simulate some diversity in the quality of the data channels (caused, e.g. by RFI), we increased noise levels in 
some of them. In particular, we increased noise from 2 to 10 times (randomly set) the standard value of 0.1\,Jy/beam in 30\% of 
the randomly chosen maps of the 18\,cm band. Similarly, in the second band (22\,cm), 10\% of maps have 2 to 5 times higher 
rms values. For the data cubes that were simulated in this way (RA, Dec, frequency), the noise levels were measured in the 
off-source region and regarded as representative of the entire channel maps.

From these measurements, the channel weights were computed according to the formulas \ref{e:newWeightingFormula} 
and \ref{e:RjFormula}, assuming $T_j$ = $D_j$ = 1. For the data and weights  
prepared in this way, RM synthesis was performed using the software written by the authors.

\begin{figure}
\centering
\includegraphics[clip, width=0.5\textwidth,clip=true]{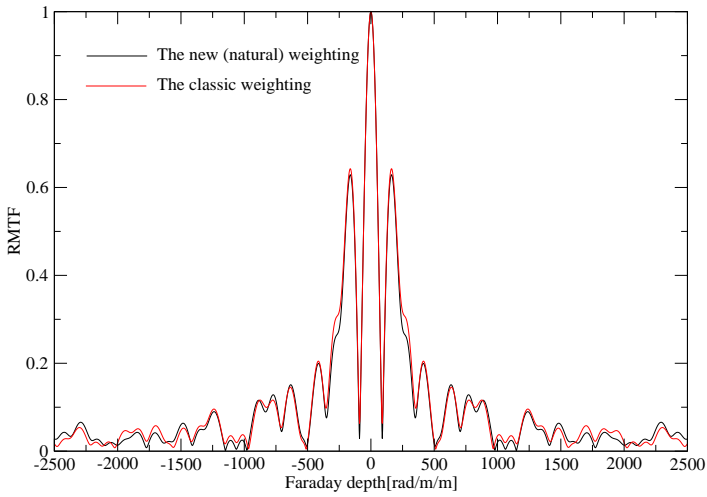}
\caption{RMTF for the new (natural) and the classic weighting obtained for the simulated dataset (see text for details).}
\label{f:RMTFNaturalClassicalSimulations}
\end{figure}

We measured noise levels (in the same way as for RA, Dec, frequency data cubes) in the real and imaginary parts of the 
obtained Faraday dispersion function (RA, Dec, Faraday depth-data cubes, Fig. \ref{f:RMQUrmsSimulations}). A comparison of 
these results with the analogous results calculated with classic (0 or 1) weighting shows that the rms values in the new 
approach are systematically lower. At the same time, the Faraday spectra obtained for the simulated sources 
(Fig. \ref{f:FaradayDispersionSimulations}) remain approximately the same for both the weightings (as expected). 
This implies that the net result of applying the new weighting is an increase of the 
signal-to-noise ratio. For these particular simulated data, the increase is about two times (the rms value is twice as low). 
We note that some small differences  seen in the presented Faraday spectrum are due to the slightly different shapes of RMTFs that were obtained 
with both  weightings (Fig. \ref{f:RMTFNaturalClassicalSimulations}). In fact, one would expect a wider main lobe of the RMTF when
calculated with the new approach, as for the radio interferometry, where usually short baselines have more weight. 
This side effect is not visible for this particular setup as here data that possesses higher rms (lower weights) were generated with nearly 
uniform distribution in the simulated spectral band. A similar approach as that presented here to
the data weighting was  
considered independently by Heald (\cite{heald08}) and Rudnick \& Owen (\cite{rudnick14}), but few details are provided there.

\subsection{Application of the new weighting scheme to the real data}
\label{s:newWeightingRealData}

\begin{figure}
\centering
\includegraphics[clip, width=0.5\textwidth,clip=true]{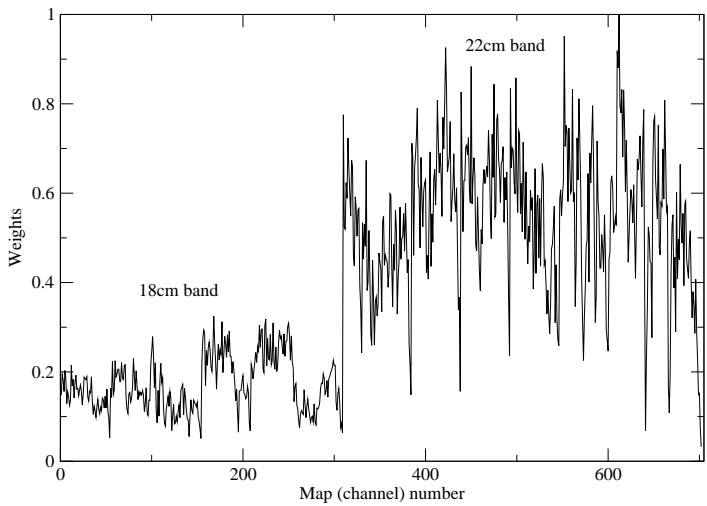}
\caption{Weights computed according to formulas \ref{e:newWeightingFormula} and \ref{e:RjFormula}, assuming $T_j$ = $D_j$ = 1 versus 
map (channel) numbers.}
\label{f:WSRTrmsWeigths}
\end{figure}

To show how the proposed new weighting scheme in the RM synthesis works in practice, we re-analysed 
WSRT-SINGS data for galaxy NGC\,2976. For comparison purposes, the dataset was processed in two ways. In the first approach, 
the classic weighting was applied with values 0 or 1. The second approach was based on the natural weighting (see above). 
Before applying RM synthesis, the data for this galaxy were re-imaged and convolved to a resolution of about 45" 
(see Sect. \ref{ss:n2976RMSynthesisResults} for a detailed description).

\begin{figure}
\centering
\includegraphics[angle=0,width=0.49\textwidth,clip=true]{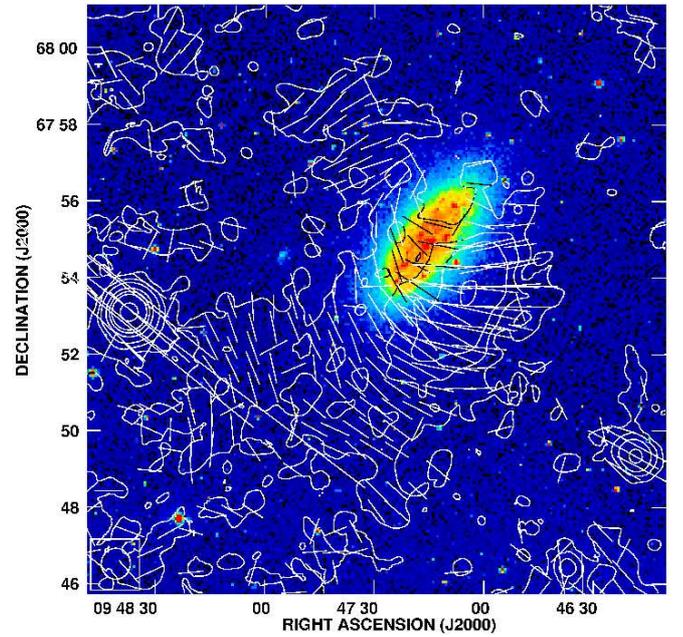}
\caption{Polarized intensity contours and apparent B-vectors (not corrected for Faraday rotation) obtained with 
the classic weighting for NGC\,2976 (WSRT-SINGS data) overlaid onto the DSS blue image. The contour levels and the beam size are the 
same as for Figure \ref{f:n2976_wsrtPI}.} 
\label{f:WSRTn2976Low}
\end{figure}

The rms noise levels measured for particular Q and U Stokes parameters channel maps were computed in a polygon area, which was free 
of polarized emission and as close as possible to the phase centre. The latter criterion 
is important to minimize the influence of a primary beam's attenuation (which is frequency-dependent) on the obtained noise measurements. 
Having measured the noise levels, it was possible to calculate the weights according to the formulas \ref{e:newWeightingFormula} 
and \ref{e:RjFormula} (Fig. \ref{f:WSRTrmsWeigths}). We note that, in the 18\,cm band, the weights are very low when compared 
to the 22\,cm band. In the next step, the RM Synthesis was performed with the obtained weights included. 
The results of applying the new weighting scheme and the classic one are compared in Figures \ref{f:n2976_wsrtPI} and \ref{f:WSRTn2976Low}, where 
distributions of polarized emission and B-vectors are shown. As  can be seen, there are less artifacts in the map obtained using the new 
weighting strategy. In fact, the noise level measured for this map is about 25\,$\mu$Jy/beam, while for the map obtained in the 
classic way, it is about 30\,$\mu$Jy/beam. Thus, because of  taking into account the appropriate weights for each frequency channel, 
it was possible to reduce noise level in the resulting PI map by about 20\%.

\subsection{Strong and weak points of the new weighting scheme}
\label{s:strongAndWeakSideOfWeighting}

Increasing sensitivity of the Faraday dispersion function is the main advantage of the new weighting scheme. 
In this method, the best quality channels have the highest weights while the poorest ones are 
underweighted. It can be easily understood  by recalling the fact that the rotation-measure synthesis is based on the Fourier transform, 
which in turn uses sine and cosine functions for a signal's sampling. The energy contained in these functions is infinite, because 
they are not limited to any particular $\lambda^2$. Thus the channels, which are less sensitive than the others, will limit sensitivity of the 
resulting power spectrum (the Faraday dispersion function). When we underweight these poor-quality channels, we  also 
automatically limit the contribution of the high level of noise associated with them (and usually also some 
imaging artifacts) to the Faraday dispersion function. We  note here that some factors, which could contribute to the reduction of 
noise level in our data, could be due to  the particular dataset's channels that were assigned higher 
weights (22\,cm)  also having had a greater field of view.

\begin{figure}
\centering
\includegraphics[clip, width=0.5\textwidth,clip=true]{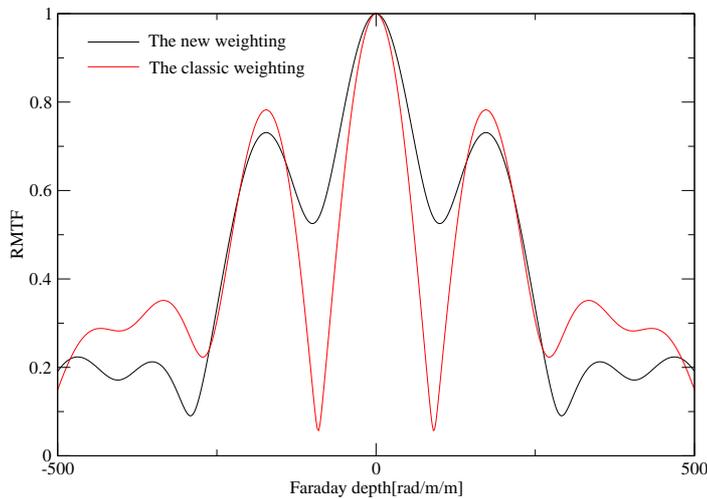}
\caption{RMTF for RM Synthesis, performed for WSRT-SINGS data of NGC\,2976 with the new and the classic weighting schemes.} 
\label{f:newWeightingRMTF}
\end{figure}

Apart from sensitivity, the new weighting can also affect  the shape of the rotation-measure transfer function. 
In Figure \ref{f:newWeightingRMTF}, both RMTFs, those obtained with the classic and the new weighting are presented. 
The latter shows a reduced level of side lobes (which is desirable) but, at the same time, their minima become much higher 
(which is not desirable). The other non-desirable effect is an increase of the width of the main lobe of RMTF in the new 
weighting (about 1.5 times), compared to the classic one.

It is also worth noting that the directions of B-vectors (not corrected for RM) on images (Fig. \ref{f:n2976_wsrtPI} and \ref{f:WSRTn2976Low}) that were obtained with 
both methods differ by  small amounts. This is due  to the mean weighted $\lambda^2$ for which directions of 
B-vectors are computed not being the same in both approaches.

\end{document}